\documentclass[
prd
,secnumarabic
,amssymb,nobibnotes, aps, prd]{revtex4}
\usepackage{amsmath}
\usepackage{graphicx}
\input{epsf}

\newcommand{\be}{\begin{equation}}    
\newcommand{\ee}{\end{equation}}
\newcommand{\beq}{\begin{eqnarray}}
\newcommand{\eeq}{\end{eqnarray}}
\newcommand{\beqn}{\begin{eqnarray*}}
\newcommand{\eeqn}{\end{eqnarray*}}

\def\nn{\nonumber}

\def\IL{\relax{\rm I\kern-.18em L}}

\begin{document}


\draft

\title{Asymptotic quasinormal modes of Reissner-Nordstr\"om and Kerr
black holes}

\author
{E. Berti and K.D. Kokkotas}
\address
{Department of Physics, Aristotle University of Thessaloniki,
Thessaloniki 54124, Greece}

\date{\today}

\begin{abstract}

According to a recent proposal, the so-called Barbero-Immirzi
parameter of Loop Quantum Gravity can be fixed, using Bohr's
correspondence principle, from a knowledge of highly-damped black hole
oscillation frequencies. Such frequencies are rather difficult to
compute, even for Schwarzschild black holes. However, it is now quite
likely that they may provide a fundamental link between classical
general relativity and quantum theories of gravity.  Here we carry out
the first numerical computation of very highly damped quasinormal
modes (QNM's) for charged and rotating black holes. In the
Reissner-Nordstr\"om case QNM frequencies and damping times show an
oscillatory behaviour as a function of charge. The oscillations become
faster as the mode order increases. At fixed mode order, QNM's
describe spirals in the complex plane as the charge is increased,
tending towards a well defined limit as the hole becomes
extremal. Kerr QNM's have a similar oscillatory behaviour when the
angular index $m=0$. For $l=m=2$ the {\it real} part of Kerr QNM
frequencies tends to $2\Omega$, $\Omega$ being the angular velocity of
the black hole horizon, while the asymptotic spacing of the {\it
imaginary} parts is given by $2\pi T_H$.

\end{abstract}


\maketitle

\narrowtext

\section{Introduction}

Quasinormal modes (QNM's) play a fundamental role in black hole
physics. They are known to ``carry the fingerprints'' of a black hole,
since their frequencies only depend on fundamental black hole
parameters such as mass, charge and angular momentum. They determine
the late-time evolution of fields in the black hole exterior. Even
more importantly, they may play a fundamental role in the newborn
field of gravitational wave astrophysics; indeed, numerical
simulations of stellar collapse and black hole collisions have shown
that in the final stage of such processes (``ringdown'') QNM's
dominate the black hole response to any kind of
perturbation. Therefore QNM's have been extensively studied for more
than thirty years (for comprehensive reviews see \cite{KS,NR}) with
the aim to shed light on the strong-field behaviour of classical
general relativity. 

Research was mainly focused on modes having small imaginary part for
two main reasons: first of all, weakly damped modes are expected to be
dominant in gravitational wave radiation; in the second place,
numerical methods used to compute QNM's generally run into trouble
when the modes' imaginary part grows, i.e., when the damping is high.
Early investigations of highly-damped modes had motivations which were
radically different from those of the present paper. Highly-damped
modes were seen as a good benchmark for the reliability of numerical
methods. Their study could provide hints (if not a formal proof) to
whether QNM's are or not infinite in number. Finally, there was hope
that their study could lead to a better understanding of the
long-standing issue of mode completeness.  Applications of different
methods yielded at first puzzling and contradictory results: WKB
methods predicted that the asymptotic real part of the frequency,
$\omega_R$, should vanish for highly damped modes \cite{GWKS}, while
the continued-fraction methods developed by Leaver \cite{L} seemed to
suggest that $\omega_R$ should be finite. An improvement of the
continued fraction technique devised by Nollert \cite{N} finally
showed that $\omega_R$ is indeed finite and determined its value;
these results were confirmed by Andersson \cite{A1} building on an
improved WKB-type technique, the phase integral method, previously
developed in \cite{AL}.

More recently QNM's, which are essentially related to the {\it
classical} dynamical properties of a black hole, have become a subject
of great interest for the quantum gravity community. This interest
stems essentially from Hod's proposal \cite{H} to apply Bohr's
correspondence principle to black hole physics. By the
Bekenstein--Hawking formula the surface area of a black hole is
nothing but its entropy. In a quantum theory of gravity the surface
area should have a discrete spectrum, and the eigenvalues of this
spectrum are likely to be uniformly spaced. In order to give a
prediction on the area spacing, Hod observed that the real parts of
the asymptotic (highly damped) quasinormal frequencies of a
Schwarzschild black hole of mass $M$, as numerically computed by
Nollert \cite{N}, can be written as

\be\label{omR}
\omega_R=T_H \ln 3. 
\ee 
where $T_H=(8\pi M)^{-1}$ is the black hole Hawking temperature (here
and in the following we use units such that $c=G=1$). He then
exploited Bohr's correspondence principle, requiring that transition
frequencies at large quantum numbers should equal classical
oscillation frequencies, to predict the spacing in the area spectrum
for a Schwarzschild black hole. It is worth stressing that in this
quantum gravity context, as opposed to the study of QNM's in the
context of gravitational wave emission, relevant modes are those for
which the imaginary part tends to infinity -- that is, modes which
damp infinitely fast and do {\it not} radiate at all.

We point out that a dynamical interpretation of the Hawking effect
through a semiclassical treatment of the quantum-mechanical
uncertainty associated to QNM oscillations was proposed long ago by
York \cite{Y}. York's dynamical treatment could successfully be used
to calculate the temperature and entropy of the hole. However, his
proposal to relate classical black hole oscillations to the hole's
quantum-mechanical behaviour is essentially different from Hod's, in
that the dynamical effects considered in \cite{Y} are dominated by
slowly damped QNM's.

Following Hod's suggestion, Bohr's correspondence principle has
recently been used by Dreyer to fix a free parameter (the so-called
Barbero-Immirzi parameter) appearing in Loop Quantum Gravity
\cite{D}. Supposing that transitions of a quantum black hole are
characterized by the appearance or disappearance of a puncture with
lowest possible spin $j_{min}$, and that changes $\Delta M$ in the
black hole mass $M$, corresponding to such a transition, are related
to the asymptotic frequency (\ref{omR}) by
\be\label{Mhom}
\Delta M=\hbar \omega_R,
\ee 

Dreyer found that Loop Quantum Gravity gives a correct prediction for
the Bekenstein-Hawking entropy if $j_{min}=1$, consequently fixing the
Barbero-Immirzi parameter. Motivated by the occurrence of an integer
value for $j_{min}$, Dreyer went on to suggest that the gauge group of
Loop Quantum Gravity should be SO(3), and not SU(2). Such a proposal
has recently been questioned in \cite{C} and \cite{Pol}. In the latter
paper, Hod's proposal has also been used as an argument in favour of
an equidistant black hole area spectrum. Evidence that the earlier
formula for black hole entropy in Loop Quantum Gravity still holds
when $j_{min}=1$ has been presented in \cite{KR}.

Numerical computations of gravitational QNM's for higher-dimensional
black holes are still lacking. However, Kunstatter \cite{K} recently
generalized Dreyer's argument to give a prediction for the area
spacing (and for the asymptotic oscillation frequency) of
$d$-dimensional black holes.

When Hod made his original proposal, the fact that $\omega_R$ is
proportional to $\ln 3$ was just a curious numerical coincidence.
Further support to the aforementioned arguments came from the
analytical proof by Motl \cite{M} that highly damped QNM frequencies
are indeed proportional to $\ln 3$.  More precisely, choosing units
such that $2M=1$, high-$n$ mode frequencies satisfy the relation
\be\label{Mresult}
\omega^{Schw}\sim{\ln 3\over 4\pi}+{i(n-1/2)\over 2}+{\cal O}(n^{-1/2}).
\ee

Such an expression has recently been confirmed through a different
analytical approach in \cite{MN}, where it has also been generalized
to higher-dimensional black holes and to four-dimensional
Reissner-Nordstr\"om (RN) black holes.  The WKB approach used in
\cite{MN} has later been exploited to analytically compute reflexion
and transmission coefficients of multidimensional Schwarzschild black
holes (and of four-dimensional RN black holes) in the limit of large
imaginary frequencies \cite{N2}.

There are many important reasons to try to understand the behaviour of
highly-damped QNM's for general black holes. Let us consider charged,
rotating black holes, having angular momentum per unit mass $a=J/M$
and charge $Q$. The black hole's (event and inner) horizons are given
in terms of the black hole parameters by
$r_\pm=M\pm\sqrt{M^2-a^2-Q^2}$. The hole's temperature
$T_H=(r_+-r_-)/A$, where $A=4\pi(2Mr_+-Q^2)$ is the hole's surface
area, related to its entropy $S$ by the relation $A=S/4$. Let us
introduce the so--called ``angular velocity of the horizon''
$\Omega\equiv 4\pi a/A$, and $\Phi\equiv 4\pi Qr_+/A$. Applying the
first law of black hole thermodynamics,
\be\label{Flaw}
\Delta M=T_H\Delta S+\Omega \Delta J+\Phi \Delta Q,
\ee
but {\it dropping without justification the $\Delta Q$ term}, and {\it
assuming that the formula for the area spectrum derived for a
Schwarzschild black hole still holds in this case}, Hod conjectured
\cite{H} that the real parts of the asymptotic frequencies for charged
and rotating black holes are given by:
\be\label{conj}
\omega_R=\tilde \omega_R\equiv T_H \ln 3+m\Omega,
\ee
where $m$ is the azimuthal eigenvalue of the field.  In particular,
such a conjecture implies that QNM's of extremal RN black holes would
have a {\it vanishing} asymptotic real part. Unfortunately, numerical
studies of asymptotic frequencies of charged and rotating black holes
have been lacking until now. Hod \cite{H2} recently used the most
systematic exploration of Kerr black hole QNM's, which was carried out
a few years ago by Onozawa \cite{O}, to lend qualitative support to
formula (\ref{conj}). However, in the following we will extend
Onozawa's numerical calculations to larger imaginary part, showing
that the use of low-order frequencies to deduce the asymptotic
behaviour as $\omega_I\to \infty$ is rather questionable.

Motl and Neitzke \cite{MN} recently obtained an analytic formula for
the asymptotic frequencies of scalar and electromagnetic-gravitational
perturbations of a RN black hole:
\be\label{MNf}
e^{\beta \omega}+2+3e^{-\beta_I \omega}=0.
\ee
For computational convenience, the authors fixed their units in a
somewhat unconventional way: they introduced a parameter $k$ related
to the black hole charge and mass by $Q/M=2\sqrt{k}/(1+k)$, so that
$\beta=4\pi/(1-k)=1/T_H$ is the inverse black hole Hawking temperature
and $\beta_I=-k^2\beta$ is the inverse Hawking temperature of the
inner horizon. However, some features of their result are particularly
puzzling:
\begin{itemize}
\item[1)]
As emphasized (and partially justified with plausibility arguments) in
\cite{MN,N2}, the predicted asymptotic RN quasinormal frequencies do
not reduce to the Schwarzschild limit as the black hole charge $Q$
tends to zero;
\item[2)]
Quasinormal frequencies of a charged black hole, according to formula
(\ref{MNf}), depend not only on the black hole's Hawking temperature,
but also on the Hawking temperature of the (causally disconnected)
inner horizon;
\item[3)]
The authors suggest that, should the black hole mass and charge
acquire a small imaginary part (which in their words ``may not be an
unreasonable thing to do'', since ``the black hole eventually
evaporates''), their asymptotic RN frequencies would be proportional
to $\ln 2$. This is in stark contrast with the Schwarzschild results:
should this be true, Dreyer's argument could be used to infer that the
gauge group of Loop Quantum Gravity is SU(2);
\item[4)]
The result does not seem to agree with the conjectured behaviour
predicted by formula (\ref{conj}). 
\end{itemize}

Therefore, their result cannot be considered conclusive, and there are
many issues to clarify. Furthermore, if asymptotic frequencies for
``generic'' black holes depend on the hole's charge and angular
momentum, a relevant question is: how should arguments based on Bohr's
correspondence principle be modified?  A similar question was recently
raised by Cardoso and Lemos \cite{CL}.  They studied the asymptotic
spectrum of Schwarzschild black holes in de Sitter spacetimes and
found that, when the black hole radius is comparable to the
cosmological radius, the asymptotic spectrum depends not only on the
hole's parameters, but also on the angular separation index $l$.

For all these reasons, a numerical computation of highly-damped QNM's
is now needed more than ever.  Such a numerical computation is
technically challenging: even though the first overtones of the QNM
spectrum have now been studied for more than 15 years, we present here
the first computation of this kind for RN and Kerr black holes.  

We will present numerical results that generally support the
calculations carried out in \cite{M,MN,N2}. Indeed, a testable
prediction of those analytical derivations is that asymptotic
frequencies for {\it scalar} perturbations should have the same value
as gravitational frequencies; however, to our knowledge, asymptotic
scalar modes have never been shown in the published literature. We
will extend Nollert's calculation to scalar modes, confirming the
analytical prediction. Furthermore, our scalar mode calculation gives
useful hints on leading order corrections to the asymptotic
frequencies of Schwarzschild black holes.

The plan of the paper is as follows. In section \ref{sec1} we describe
our numerical method, extending Nollert's technique to RN and Kerr
black holes. In section \ref{sec2} we show our numerical results, and
put forward some conjectures on their implications for the asymptotic
behaviour of the modes. The conclusions and a discussion follow.

\section{Numerical method}\label{sec1}

A first comprehensive analysis of the QNM spectrum of RN black holes
was first carried out by Gunter \cite{G}; then Kokkotas and Schutz
\cite{KSc} verified and extended his results using numerical
integrations and WKB methods. Unfortunately, the standard WKB
techniques cannot be applied to our case, since they become inaccurate
in estimating the modes' real part as the imaginary part increases,
unless one resorts to more sophisticated phase-integral methods
\cite{AL}. The first few modes of the Schwarzschild and Kerr black
holes were studied by Leaver using a continued fraction technique
\cite{L}, which was then extended to the RN case \cite{L2}.  This
technique is generally rather accurate for modes having $\omega_I\sim
\omega_R$, but it eventually loses accuracy when $\omega_I\gg
\omega_R$. The error is essentially introduced by a truncation of the
power-series solution to the radial equation at some large $N$.  For
large $N$, Leaver has shown that this error can be written as an
integral, which is rapidly convergent near the lower quasinormal
frequencies (where $|\omega_R|>|\omega_I|$ and $|\omega|\sim 1$);
however, the convergence becomes slower as the overtone index
increases \cite{L2}.  That's why the problem of computing high-order
overtones is so numerically challenging. An improvement of the
continued fraction method was used by Nollert \cite{N} to find the
asymptotic behaviour of the QNM frequencies for Schwarzschild black
holes. In the following sections we show how to generalize Nollert's
method to the charged and rotating cases.

\subsection{Reissner-Nordstr\"om black holes}

In this section and the following we briefly describe the
computational procedure we used. More details are given, eg., in
\cite{N,L,L2}. Let us introduce a tortoise coordinate $r_*$, defined
in the usual way by the relation
\be
{d r\over d r_*}={\Delta\over r^2},
\ee
where $\Delta=r^2-2Mr+Q^2$; after a separation of the angular
dependence and a Fourier decomposition, axial electromagnetic and
gravitational perturbations of a RN metric are described by a couple
of wave equations:
\be\label{axial}
\left({d^2\over dr_*^2}+\omega^2\right)Z_i^-=V_i^-Z_i^-.
\ee
Polar perturbations can be obtained from the axial ones through a
Chandrasekhar transformation \cite{MTB}.  In the limit $Q=0$, the
potentials $V_1^-$ and $V_2^-$ describe, respectively, purely
electromagnetic and axial gravitational perturbations of a
Schwarzschild black hole.  Now, the radial equations for the
perturbations can be solved using a series expansion around some
suitably chosen point. The coefficients $a_n$ of the expansion are
then determined by a recursion relation. For Schwarzschild black holes
the recursion relation has three terms, i.e., it is of the form:
\beq
&&\alpha_0 a_1+\beta_0 a_0=0,\\
&&\alpha_1 a_2+\beta_1 a_1+\gamma_1 a_0=0,\nn\\
&&\alpha_n a_{n+1}+\beta_n a_n+\gamma_n a_{n-1}+\delta_n a_{n-2}=0,
\qquad n=2,3,\dots\nn
\eeq
where the recursion coefficients $\alpha_n$, $\beta_n$ and $\gamma_n$
are functions of the frequency $\omega$ and of $l$ (if we fix units
such that $2M=1$).

For RN black holes we actually have a four-term recursion relation
(whose coefficients also depend on the charge $Q$), but we can reduce
it to the previous form using a Gaussian elimination step. It turns
out that the QNM boundary conditions are satisfied when the following
continued-fraction condition on the recursion coefficients holds:
\be\label{CF}
0=\beta_0-
{\alpha_0\gamma_1\over \beta_1-}
{\alpha_1\gamma_2\over \beta_2-}\dots
\ee
The $n$--th quasinormal frequency is (numerically) the most stable
root of the $n$--th inversion of the continued-fraction relation
(\ref{CF}), i.e., it is the root of
\beq\label{CFI}
&&\beta_n-
{\alpha_{n-1}\gamma_{n}\over \beta_{n-1}-}
{\alpha_{n-2}\gamma_{n-1}\over \beta_{n-2}-}\dots
{\alpha_{0}\gamma_{1}\over \beta_{0}}
=\\
&=&{\alpha_n\gamma_{n+1}\over \beta_{n+1}-}
{\alpha_{n+1}\gamma_{n+2}\over \beta_{n+2}-}\dots
\qquad (n=1,2,\dots).\nn
\eeq
The infinite continued fraction appearing in equation (\ref{CFI}) can
be summed ``bottom to top'' starting from some large truncation index
$N$. Nollert has shown that the convergence of the procedure improves
if such a sum is started using a wise choice for the value of the
``rest'' of the continued fraction, $R_N$. This rest can be obtained
solving the equation
\be
R_N={\gamma_{N+1}\over\beta_{N+1}-\alpha_{N+1}R_{N+1}},
\ee
and assuming that
\be\label{RN}
R_N=\sum_{k=0}^{\infty}C_k N^{-k/2}.
\ee
If we introduce $\rho\equiv-i\omega$, and denote by $r_+$ the
coordinate radius of the black hole's event horizon, it turns out that
the first few coefficients in the series are $C_0=-1$,
$C_1=\pm\sqrt{2\rho (2r_+-1)}$, $C_2=(3/4-2\rho r_+)$.

As first shown in \cite{A2} using WKB techniques and then confirmed
using different numerical methods in \cite{AO}, even for modes with
moderate imaginary parts, RN QNM frequencies show a very peculiar
behaviour as the charge increases. Our numerical results agree with
those shown in \cite{AO}.  Furthermore, we have checked that our
method yields Nollert's asymptotic frequencies in the Schwarzschild
limit.  Due to convergence reasons, calculations become more and more
computationally intensive as the charge is increased. Indeed, because
of the merging of the inner and outer horizons, maximally charged
black holes yield radial equations for the perturbation variables
which have a different singularity structure, and deserve a special
treatment \cite{OMOI}. We plan to investigate the behaviour of highly
damped QNM's for maximally charged black holes in the future.

\subsection{Kerr black holes}

As we did for the charged black hole case, here we only briefly sketch
our computational procedure, referring the reader to \cite{N,L,O} for
more details.  In the Kerr case, the perturbation problem reduces to a
couple of differential equations - one for the angular part of the
perturbations, and the other for the radial part.  In Boyer-Lindquist
coordinates, defining $u=\cos\theta$, the angular equation reads
\be
\left[
(1-u^2)S_{lm,u}
\right]_{,u}
+\left[
(a\omega u)^2-2a\omega su+s+A_{lm}-{(m+su)^2\over 1-u^2}
\right]
S_{lm}=0,
\ee
and the radial one is
\be
\Delta R_{lm,rr}+(s+1)(2r-1)R_{lm,r}+V(r)R_{lm}=0,
\ee
where
\beq
V(r)&=&\left\{
\left[
(r^2+a^2)^2\omega^2-2am\omega r+a^2m^2+is(am(2r-1)-\omega(r^2-a^2))
\right]\Delta^{-1}\right.\nn
\\
&+&
\left.
\left[
2is\omega r-a^2\omega^2-A_{lm}
\right]
\right\}.
\eeq

In writing down these equation we have adopted, consistently with what
we did for the RN case, Leaver's conventions. In particular, we chose
units such that $2M=1$. The parameter $s=0,-1,-2$ for scalar,
electromagnetic and gravitational perturbations respectively, $a$ is
the Kerr rotation parameter ($0<a<1/2$), and $A_{lm}$ is an angular
separation constant. In the Schwarzschild limit the angular
separation constant can be determined analytically, and is given by
the relation $A_{lm}=l(l+1)-s(s+1)$.

Boundary conditions for each equation translate into a couple of
three-term continued fraction relations of the form (\ref{CF}).
Finding QNM frequencies is now a two-step procedure: for assigned
values of $a,~\ell,~m$ and $\omega$, first find the angular separation
constant $A_{lm}(\omega)$ looking for zeros of the {\it angular}
continued fraction; then replace the corresponding eigenvalue into the
{\it radial} continued fraction, and look for its zeros as a function
of $\omega$.  In principle, the convergence of the procedure for modes
with large imaginary parts can be improved, as described earlier, by a
wise choice of the rest, $R_N$, of the radial continued
fraction. Expanding this rest as in formula (\ref{RN}) and introducing
$b\equiv \sqrt{1-4a^2}$, we get for the first few coefficients:
$C_0=-1$, $C_1=\pm\sqrt{2\rho b}$, $C_2=\left[ 3/4-\rho (b+1)-s
\right]$.

As for the RN case, we have checked that our results agree with those
shown in \cite{L,O} for small values of $\omega_I$, and that we get
Nollert's asymptotic quasinormal frequencies in the non rotating limit.

\section{Results}\label{sec2}

As a first step in the computation of asymptotic modes for charged and
rotating black holes we have checked that our numerical methods
reproduce known results in the Schwarzschild limit \cite{units}. In
particular, we have verified the asymptotic behaviour found by Nollert
\cite{N}. His main result was that the real part of the asymptotic
QNM frequencies corresponding to {\it gravitational} perturbations can
be well fitted by a relation of the form
\be\label{fit}
\omega_R=\omega_{\infty}+{\lambda_{s,l}\over \sqrt{n}}.
\ee
The leading-order fitting coefficient is independent of $l$ and given
by $\omega_{\infty}=0.0874247$, consistently (within numerical
accuracy) with the analytical formula (\ref{Mresult}).  Corrections of
order $\sim n^{-1/2}$, however, are $l$-dependent. Furthermore, we
will see in a moment that they also depend on the spin $s$ of the
perturbing field, and that's why we denoted them by
$\lambda_{s,l}$. For gravitational perturbations ($s=-2$) Nollert
found (and we verified to the same level of accuracy) that
$\lambda_{-2,2}=0.4850$, $\lambda_{-2,3}=1.067$,
$\lambda_{-2,6}=3.97$.

Now, an important testable prediction of the recent derivations of the
``$\ln 3$'' asymptotic behaviour is that {\it scalar} black hole
perturbations should lead to the same asymptotic QNM frequency, at
leading order in an expansion in powers of $n^{-1/2}$ \cite{M,MN,N2}.
To check this prediction we have extended our numerical calculations
to $s=0$. Using the procedure described in \cite{N}, we found that our
scalar QNM data are well fitted by formula (\ref{fit}). As $n\to
\infty$, the asymptotic frequency for scalar modes is again given by
$\omega_{\infty}=0.0874247$, consistently with the analytic
calculation. What changes are the numerical values of the
leading-order correction coefficients $\lambda_{0,l}$. Namely, we
find: $\lambda_{0,0}=0.0970$, $\lambda_{0,1}=0.679$,
$\lambda_{0,2}=1.85$.

Neitzke \cite{N2} recently suggested that leading order corrections to
the asymptotic frequency should be proportional to
$[(s^2-1)-3l(l+1)]$.  As an interesting by-product of our calculation,
we found that our numerical values for the $\lambda_{s,l}$'s are
consistent with Neitzke's conjecture; that the proportionality
constant is dependent on $s$,
\be\label{fit2}
\lambda_{s,l}=k_s[(s^2-1)-3l(l+1)];
\ee
and finally, we determined the proportionality constants $k_s$ to be
given by $k_{-2}=-0.0323$, $k_{0}=-0.0970$.  Recently Maassen van den
Brink \cite{MVDB} derived $k_{-2}$ analytically, finding
\be\label{km2}
k_{-2}=-{\sqrt{2}\Gamma(1/4)^4\over 432 \pi^{5/2}}\simeq -0.0323356.
\ee
Notice that, within our numerical accuracy, $k_{0}=3k_{-2}$. This
prediction may be helpful to analytically determine $k_0$.  We shall
say more about the importance of leading-order corrections to the
frequencies in the following; now we turn to a discussion of our
numerical results for RN and Kerr black holes.

\subsection{Reissner-Nordstr\"om black holes}

\begin{figure}[h]
\centering
\includegraphics[angle=270,width=8cm,clip]{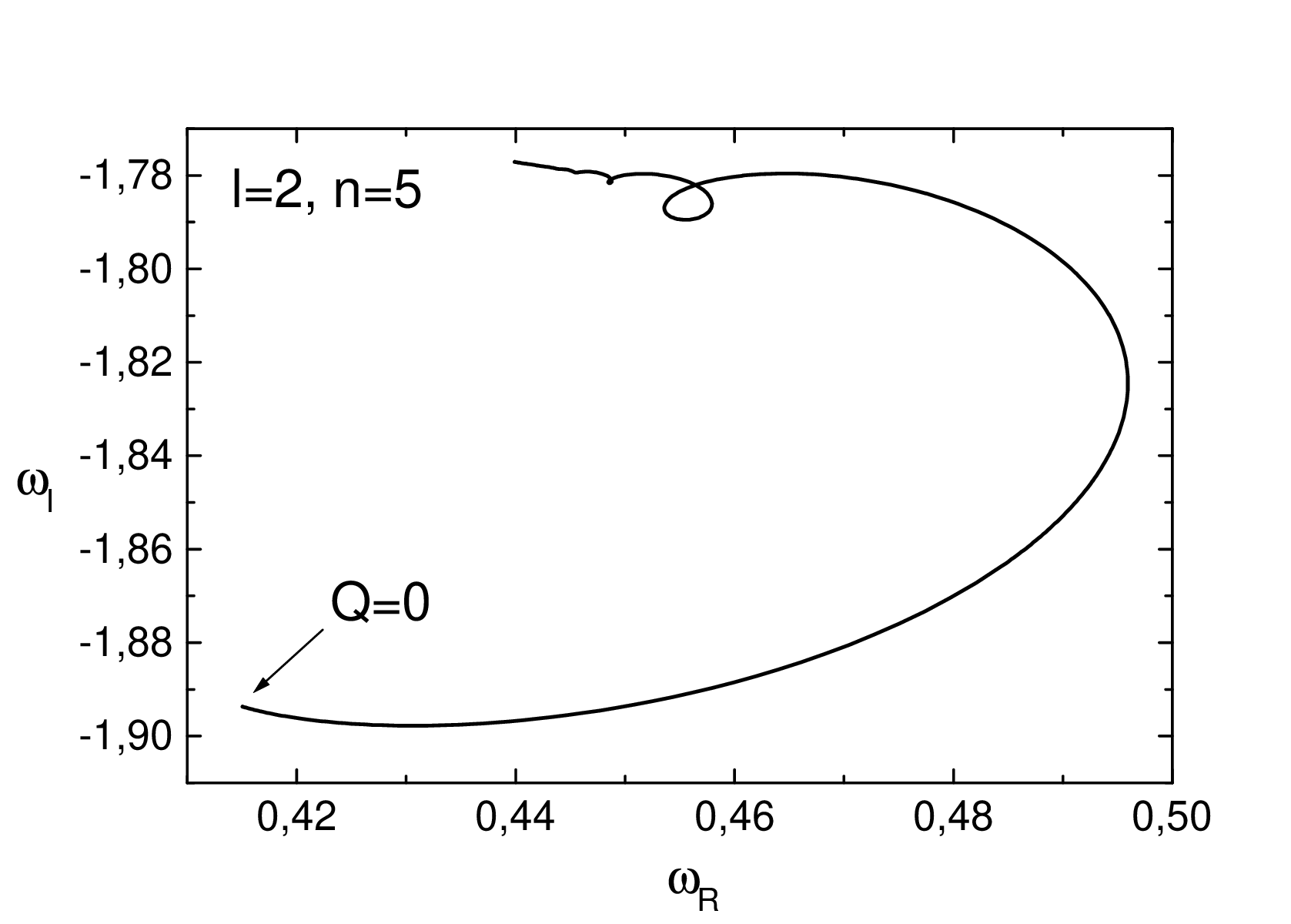}
\includegraphics[angle=270,width=8cm,clip]{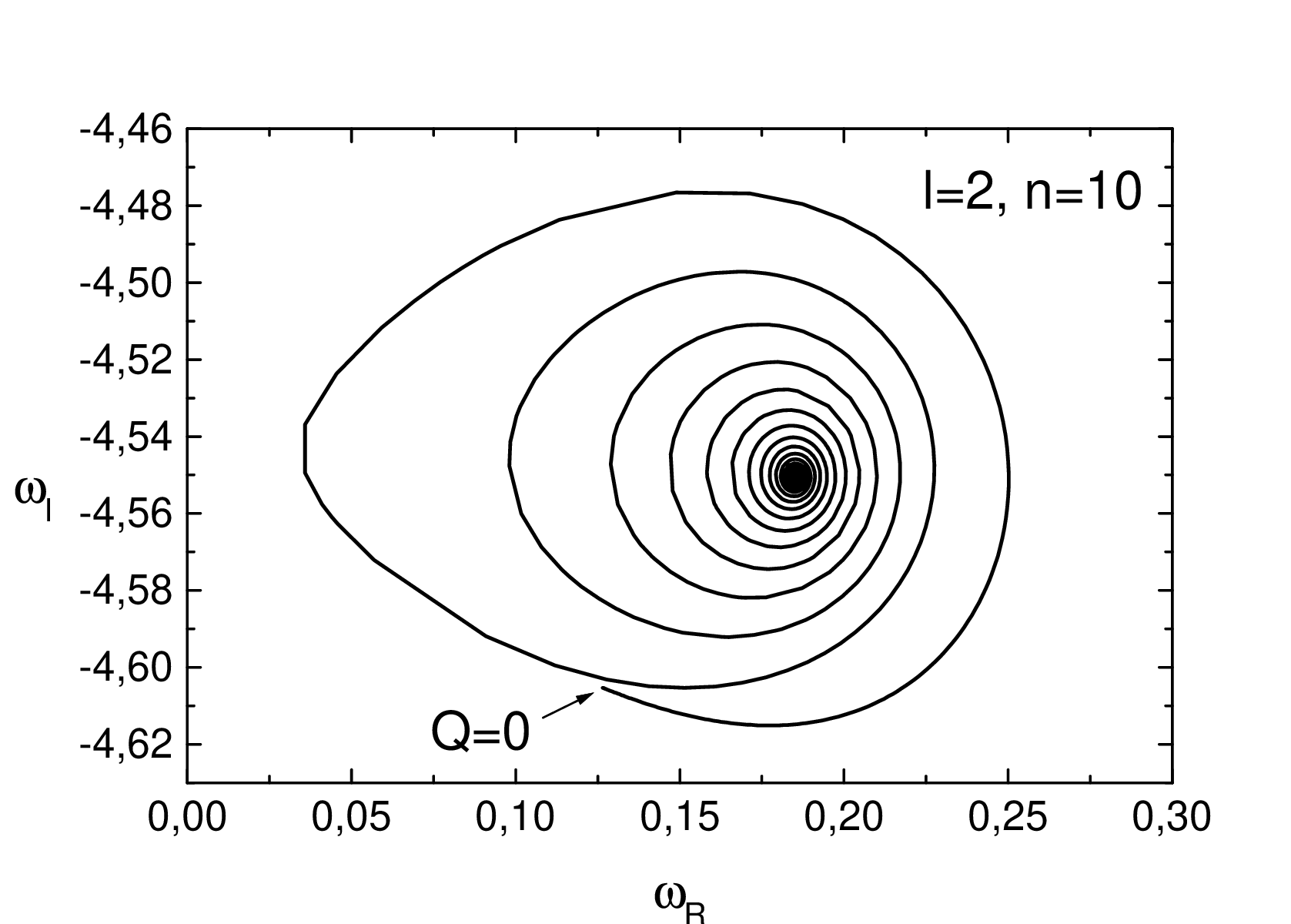}
\includegraphics[angle=270,width=8cm,clip]{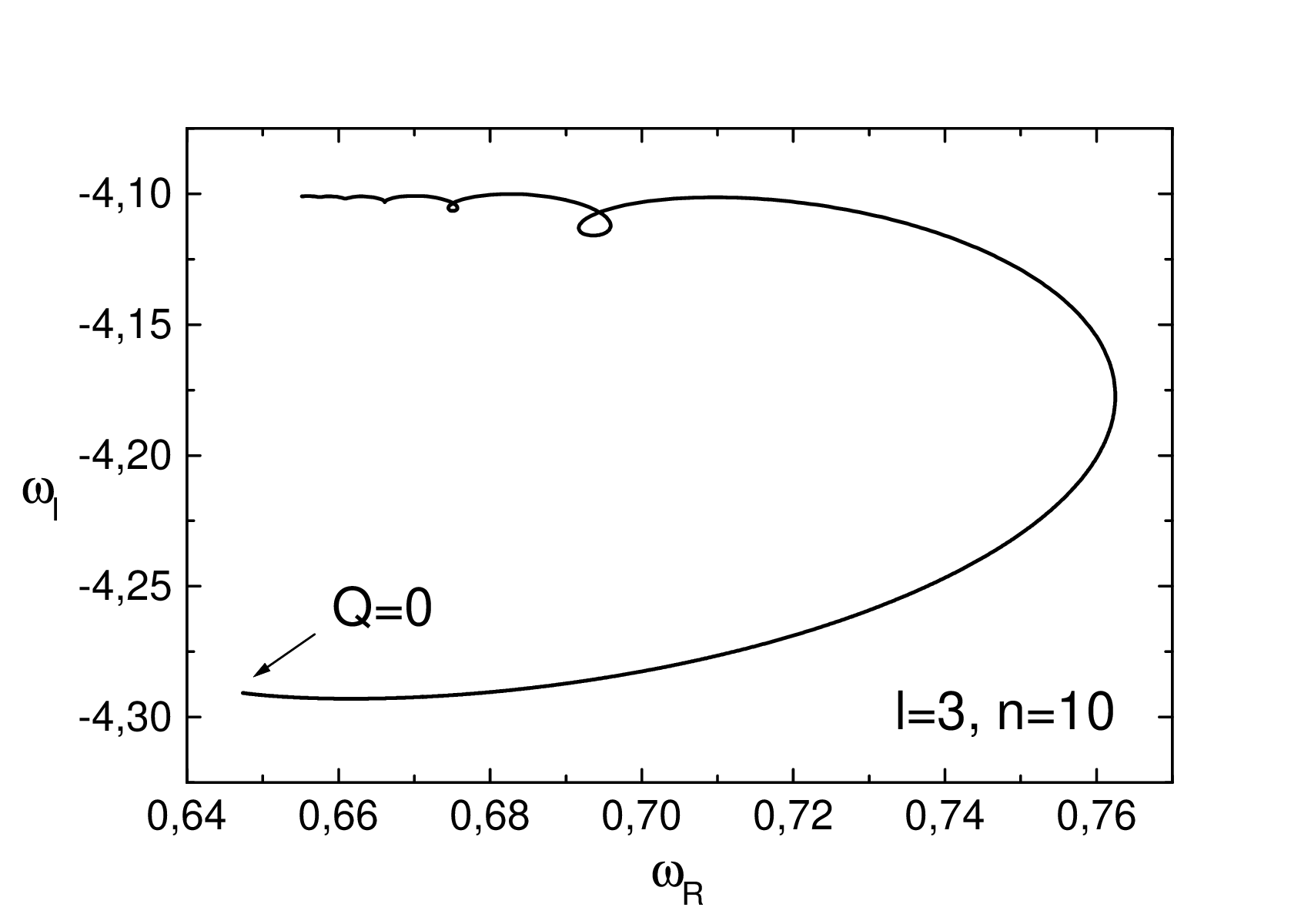}
\includegraphics[angle=270,width=8cm,clip]{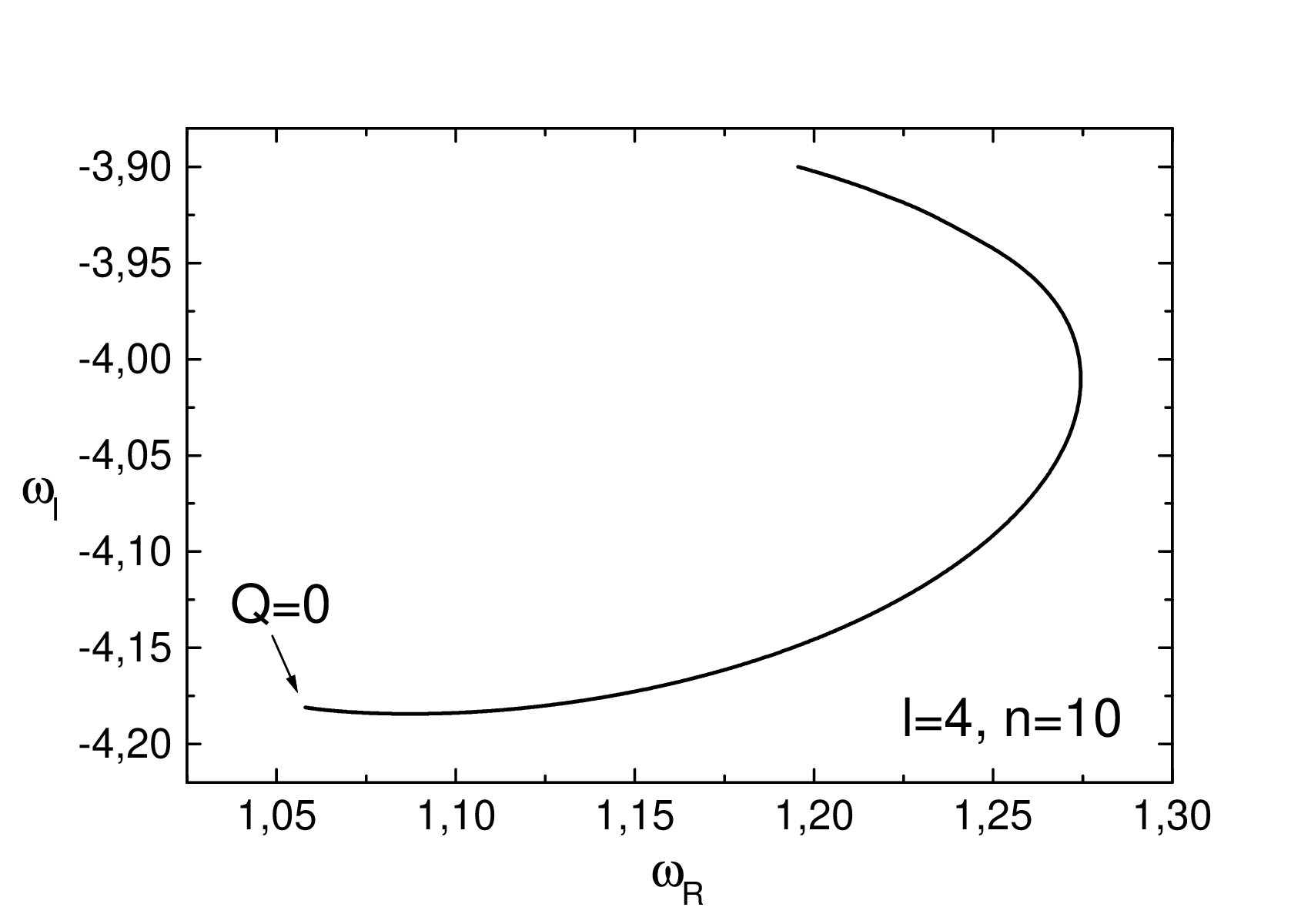}
\includegraphics[angle=270,width=8cm,clip]{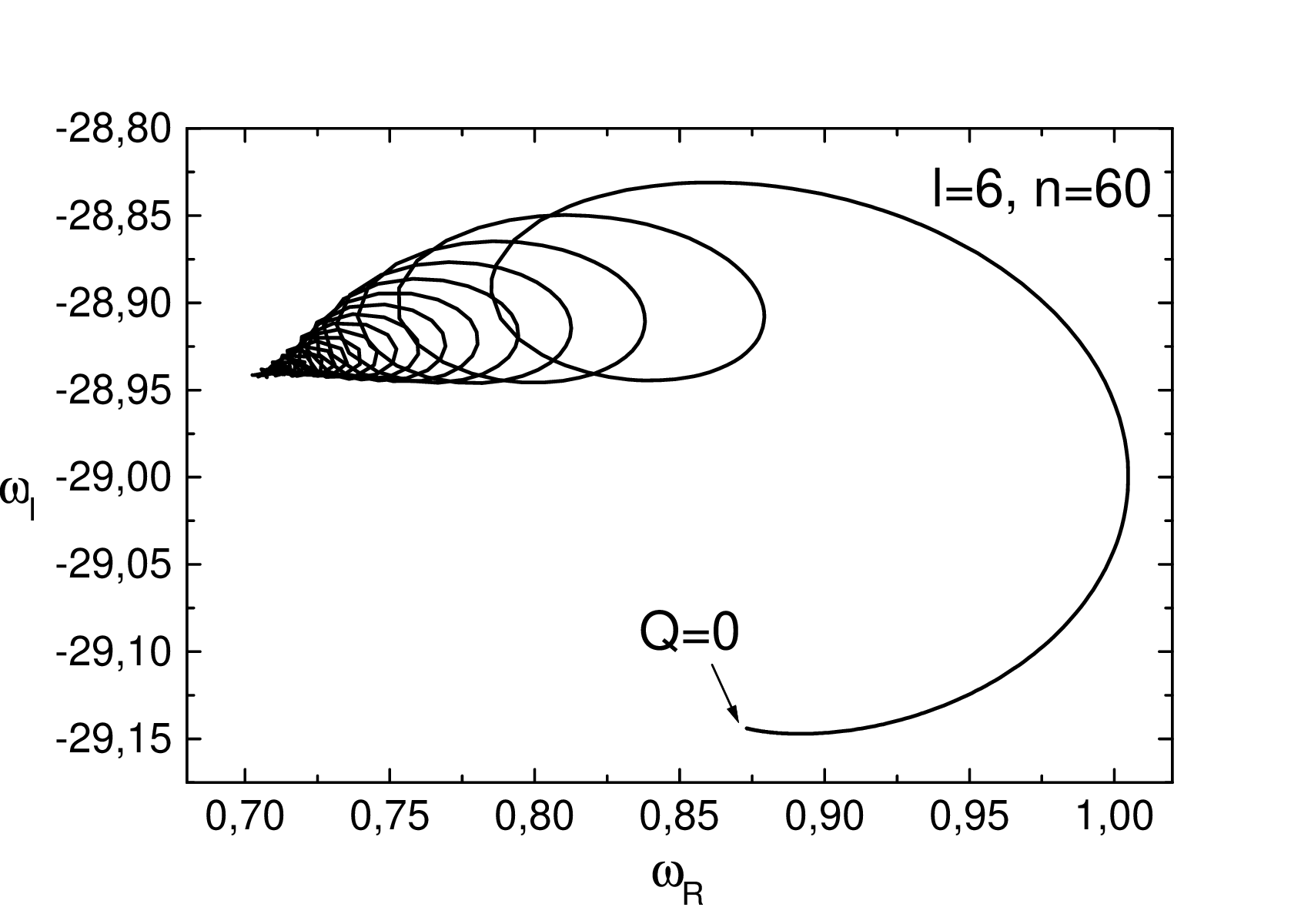}
\caption{
The top two panels show the behaviour of the $n=5$ and $n=10$ QNM
frequencies in the complex $\omega$ plane. The $n=10$ mode ``spirals
in'' towards its value in the extremal charge limit; the number of
spirals described by each mode increases roughly as the mode order
$n$. The panels in the second row show how the $n=10$ spiral
``unwinds'' as the angular index $l$ is increased (in other words, the
asymptotic behaviour sets in later for larger $l$'s). Finally, the
bottom panel shows a high-$l$ mode trajectory ``pointing'' to its
limit as the charge becomes extremal. In all cases, we have marked by
an arrow the frequency corresponding to the Schwarzschild limit
($Q=0$).
}
\label{fig1}
\end{figure}

The numerical behaviour of the first few overtones of a RN black hole
was studied numerically by Andersson and Onozawa \cite{AO}, who
unveiled a very peculiar behaviour as the modes' imaginary part
increases. Our numerical codes are in excellent agreement with their
results. In comparing with their paper note, however, that Andersson
and Onozawa count modes starting from $n=0$, while we label the
fundamental mode by $n=1$, following Leaver (see table 1 in
\cite{L}). The results we display refer to perturbations reducing to
pure gravitational perturbations of Schwarzschild in the uncharged
limit. The trajectories described by the modes in the complex-$\omega$
plane first show ``closed loops'', as in the top left panel of figure
\ref{fig1}. Then they get a spiral-like shape, moving out of their
Schwarzschild value and ``looping in'' towards some limiting frequency
as $Q$ tends to the extremal value. This kind of behaviour is shown in
the top right panel of figure \ref{fig1}. We have observed that such a
spiralling behaviour sets in for larger values of the modes' imaginary
part (i.e., larger values of $n$) as the angular index $l$
increases. In other words, increasing $l$ for a given value of the
mode index $n$ has the effect of ``unwinding'' the spirals, as can be
observed in the two panels in the second row of figure
\ref{fig1}. However, for each $l$ the spiralling behaviour is
eventually observed when $n$ is large enough; a typical example for
$l=6$ is shown in the last plot of figure \ref{fig1}.

\begin{figure}[h]
\centering
{\includegraphics[angle=270,width=8cm,clip]{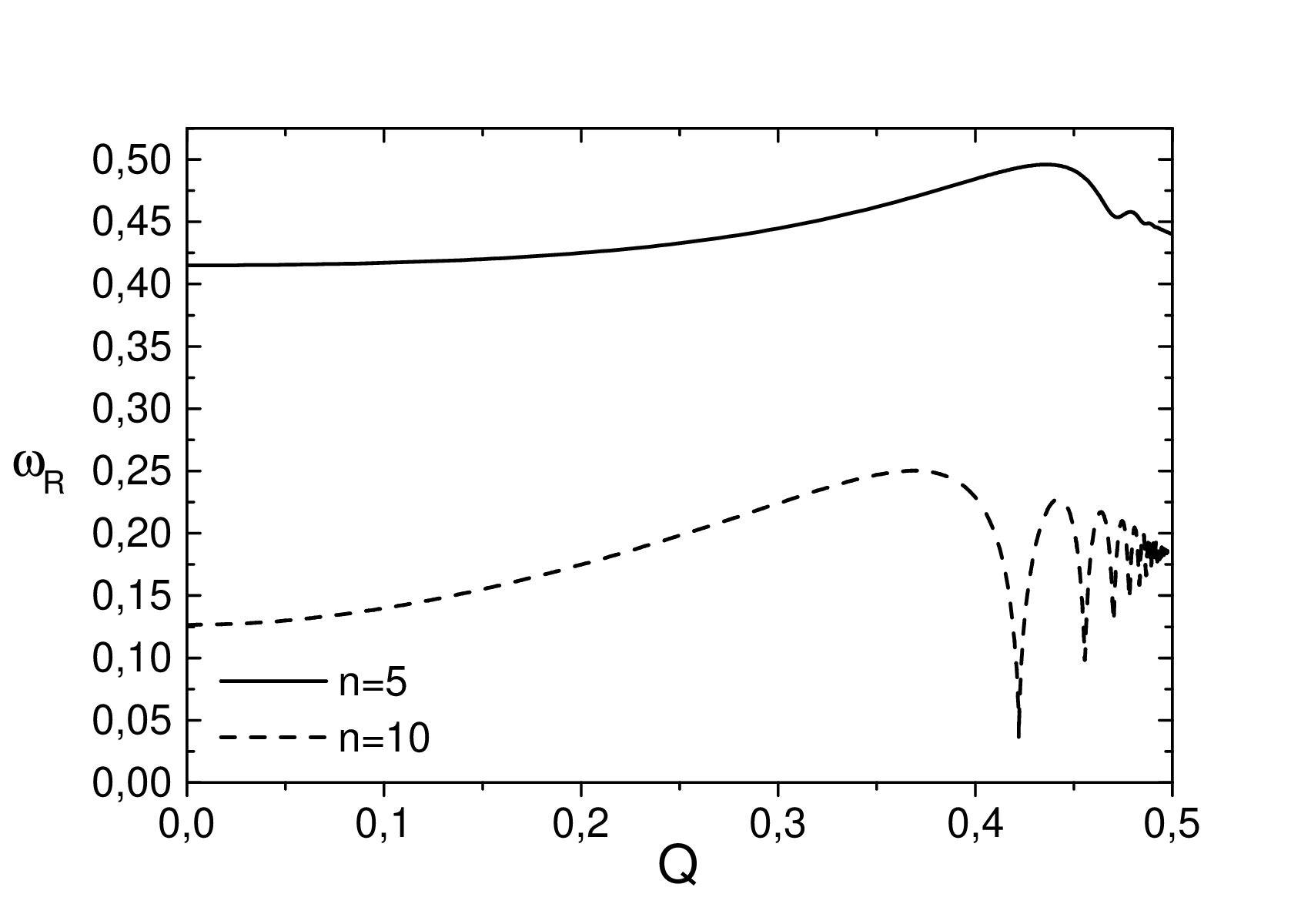}
\includegraphics[angle=270,width=8cm,clip]{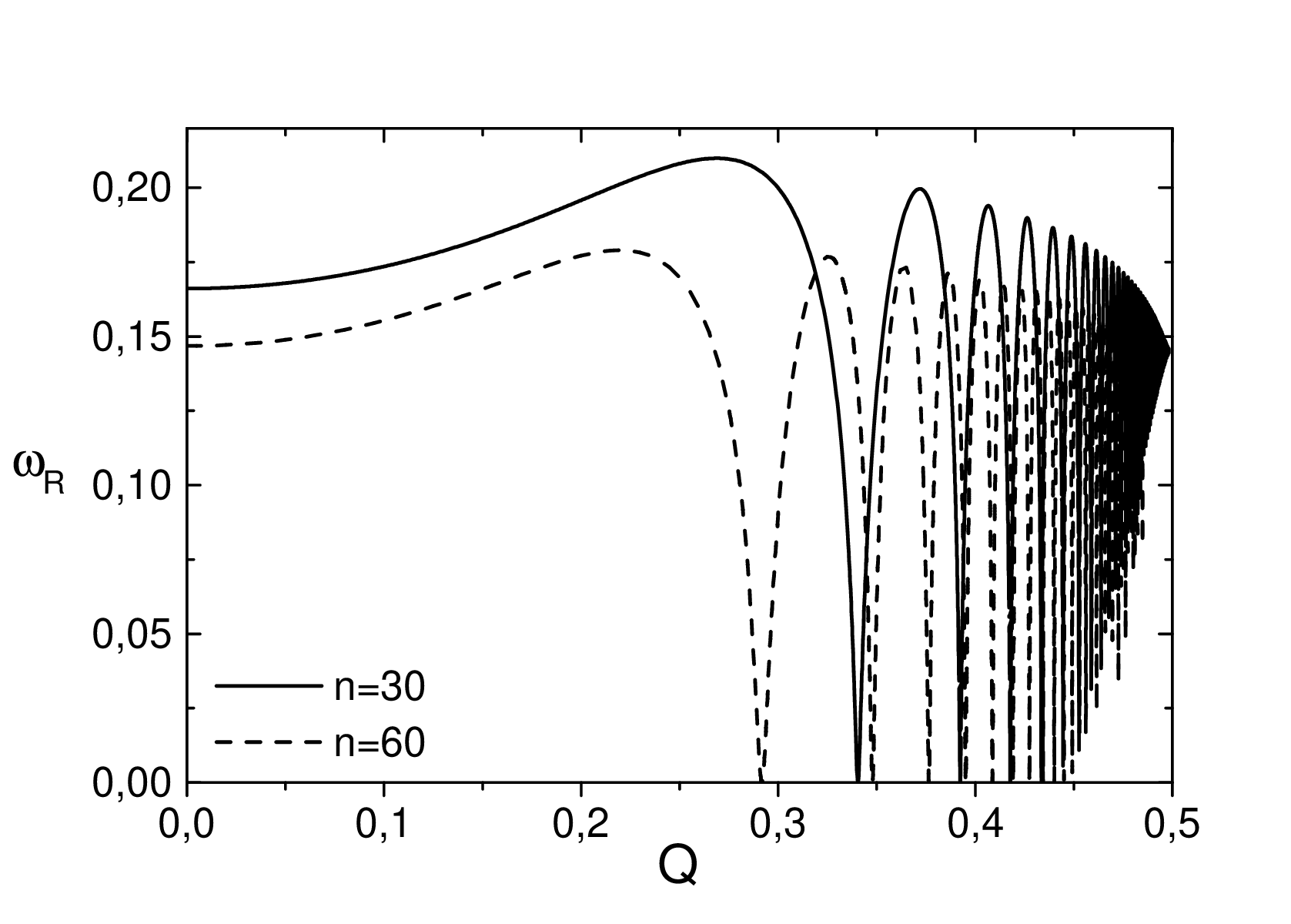}}
{\includegraphics[angle=270,width=8cm,clip]{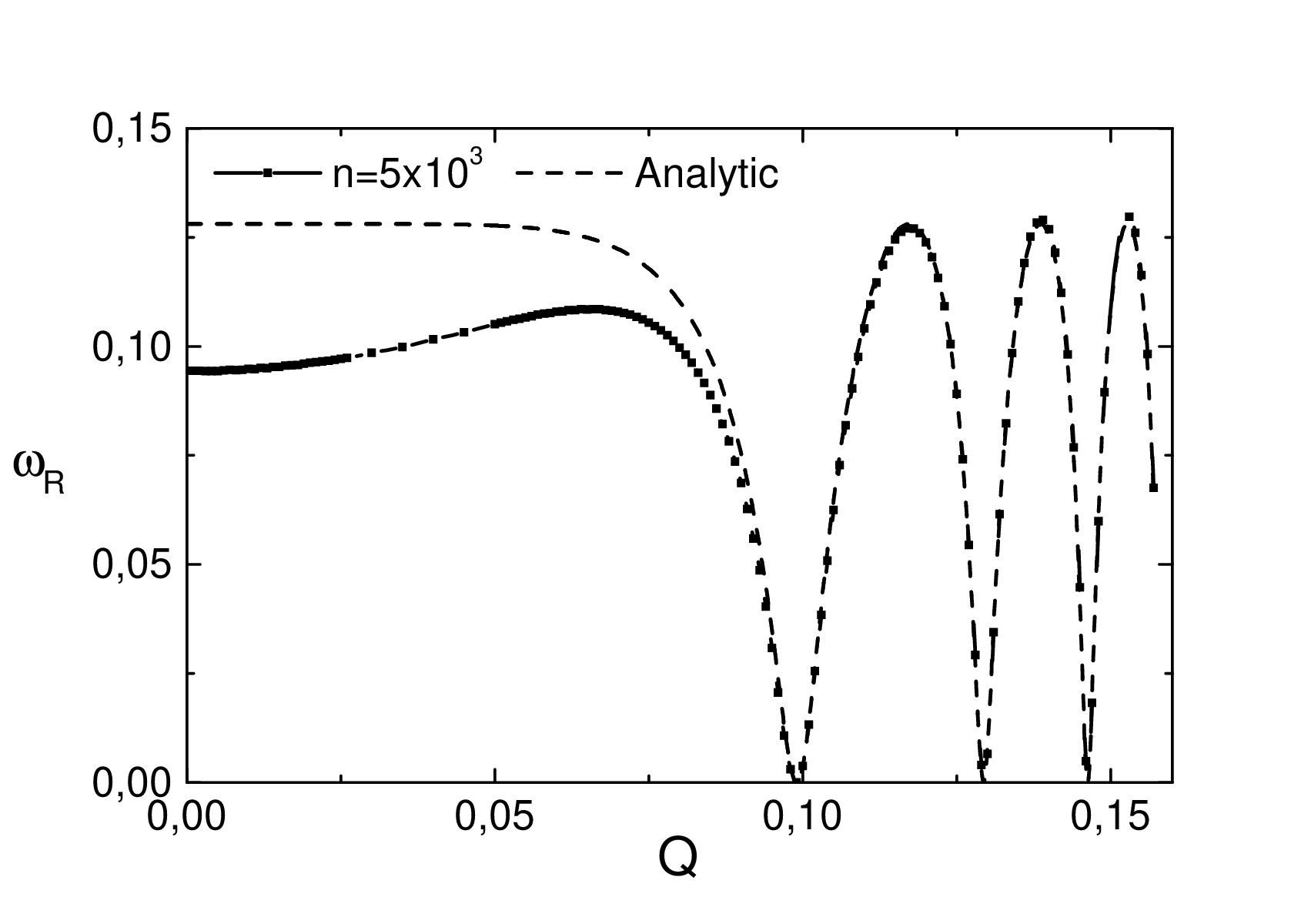}
\includegraphics[angle=270,width=8cm,clip]{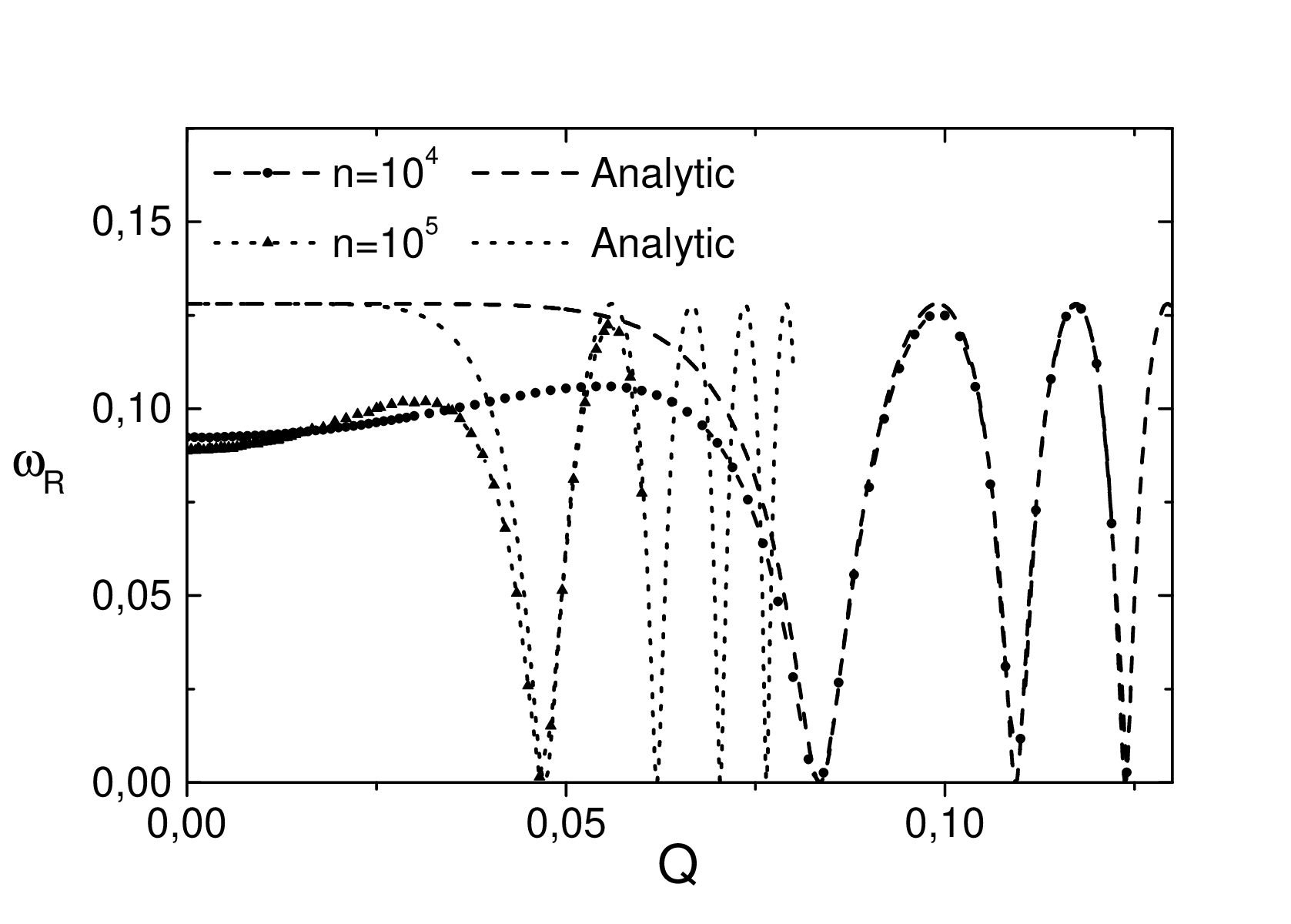}}
\caption{
Real part of the RN QNM frequencies as a function of charge for
$n=5,~10,~30,~60,~5000,~10000,~100000$. As the mode order increases
the computation becomes more and more time consuming, the oscillations
become faster, and a good numerical sampling is rather difficult to
achieve; therefore in the last plot we use different symbols (small
squares, circles and triangles) to display the actually computed
points. For $n=5000,~10000,~100000$ we also compare to the prediction
of the analytic formula (\ref{MNf}) derived by Motl and Neitzke
\cite{MN}. The oscillatory behaviour is reproduced extremely well by
their formula, but the disagreement increases for small charge:
formula (\ref{MNf}) does not yield the correct Schwarzschild limit.
}
\label{fig2}
\end{figure}

Maybe a clearer picture of the modes' behaviour can be obtained
looking separately at the real and imaginary parts of the mode
frequencies as function of charge. Let us focus first on the real
parts. The corresponding numerical results are shown in figure
\ref{fig2}, together with the predictions of the analytic formula
(\ref{MNf}).  It's quite apparent that, as the mode order grows, the
oscillating behaviour as a function of charge start earlier and
earlier. As $n$ increases, the oscillations become faster, the
convergence of the continued fraction method slower, and the required
computing time gets longer. Therefore, when the imaginary part
increases it becomes more difficult to follow the roots numerically as
we approach the extremal value $Q=1/2$. That's why our data for large
values of $n$ do not cover the whole range of allowed values for
$Q$. Despite these difficulties, we have many reasons to trust our
numerics. We have carefully checked our results, using first double
and then quadrupole precision in our Fortran codes (indeed, as $n$
increases, we can obtain results for large values of the charge only
using quadrupole precision). As we have shown earlier, our frequencies
accurately reproduce Nollert's results in the Schwarzschild limit, so
our numerics can be trusted for small values of $Q$.  Furthermore, the
predictions of the analytic formula exactly overlap with the
oscillations we observe for large values of the charge. Not only this
does give support to the asymptotic formula \cite{N2}; it also gives
us faith that the numerics are meaningful also for large charge, where
we don't have any published results to confirm our predictions.

\begin{figure}[h]
\centering
\includegraphics[angle=270,width=8cm,clip]{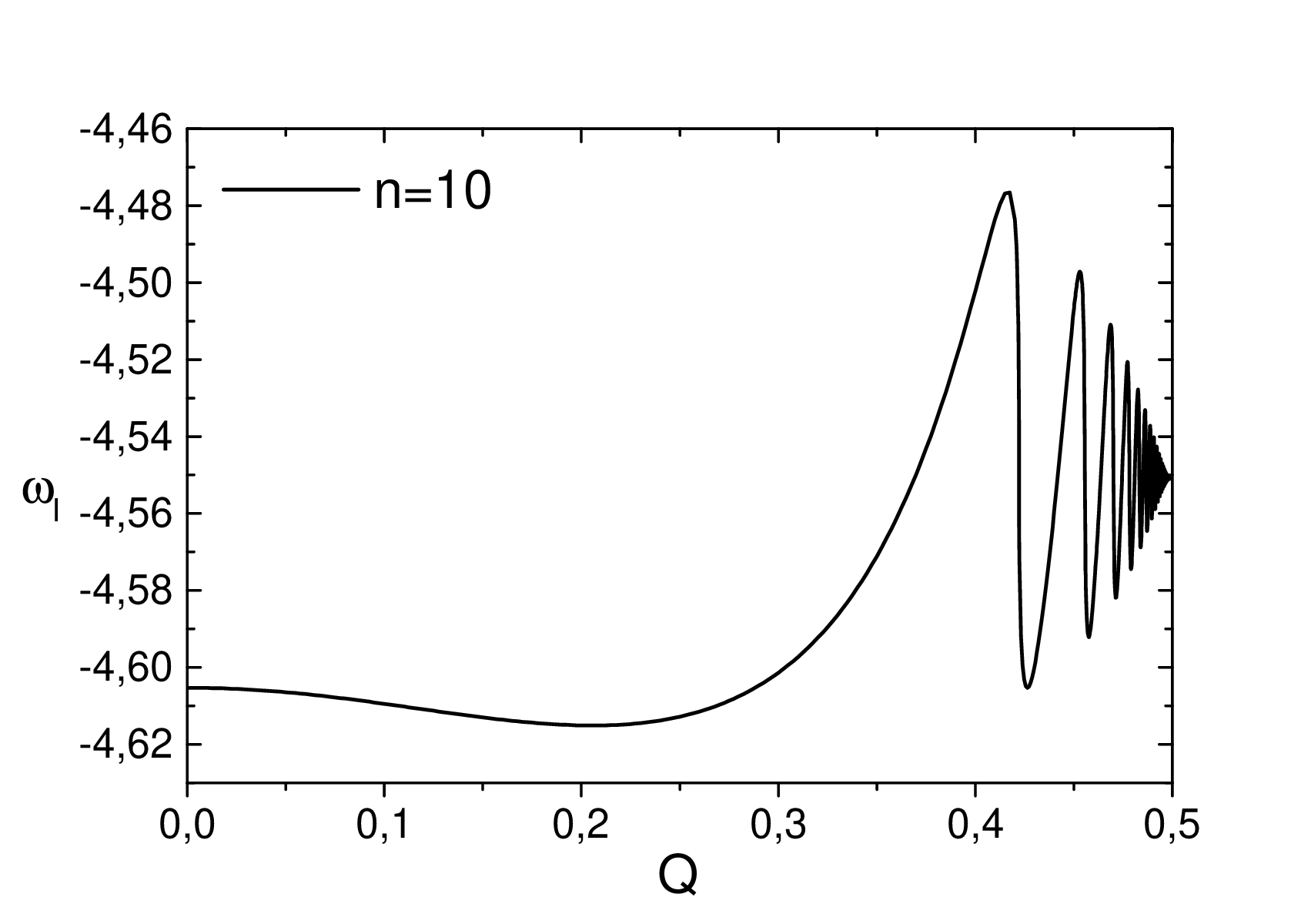}
\includegraphics[angle=270,width=8cm,clip]{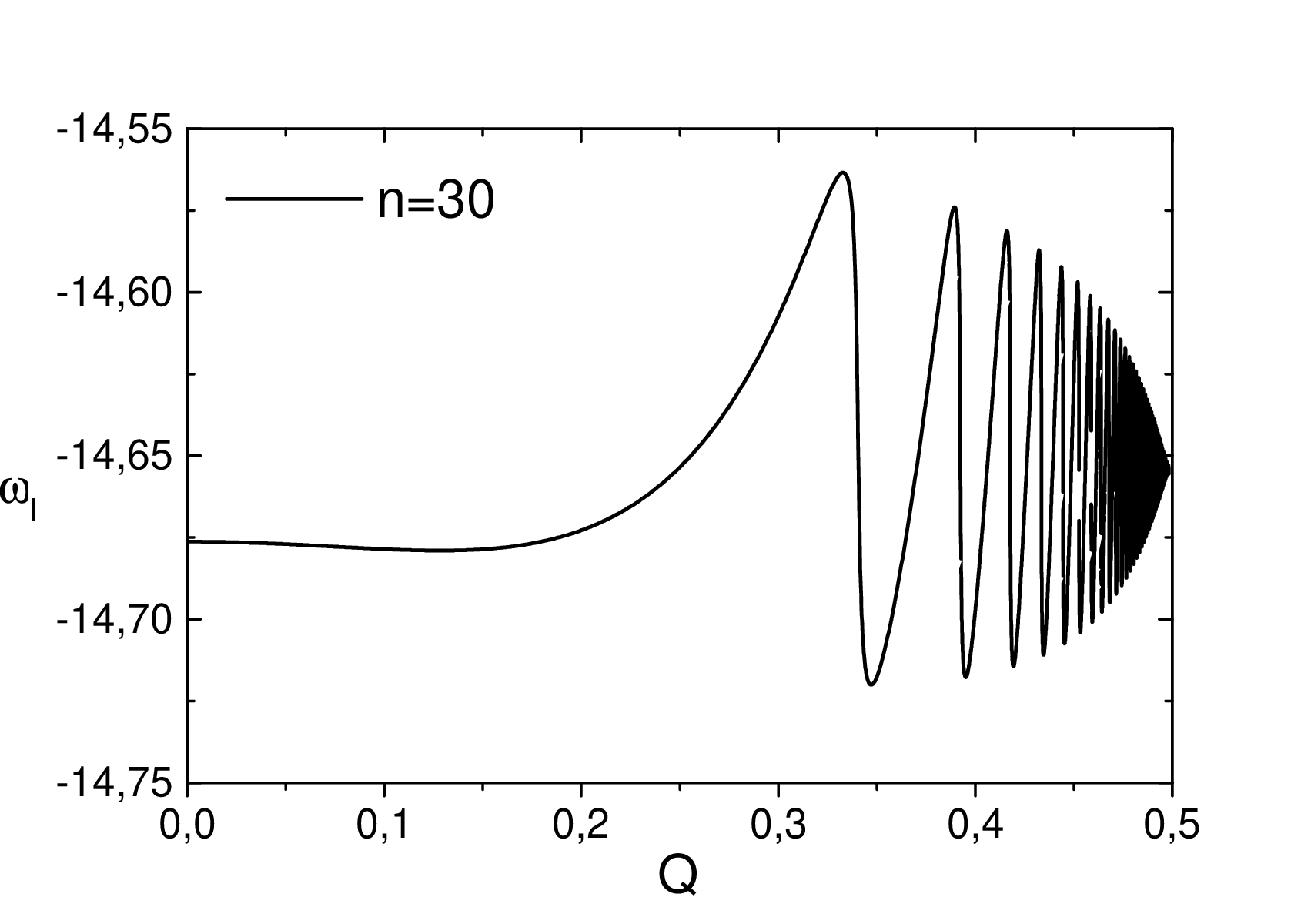}
\includegraphics[angle=270,width=8cm,clip]{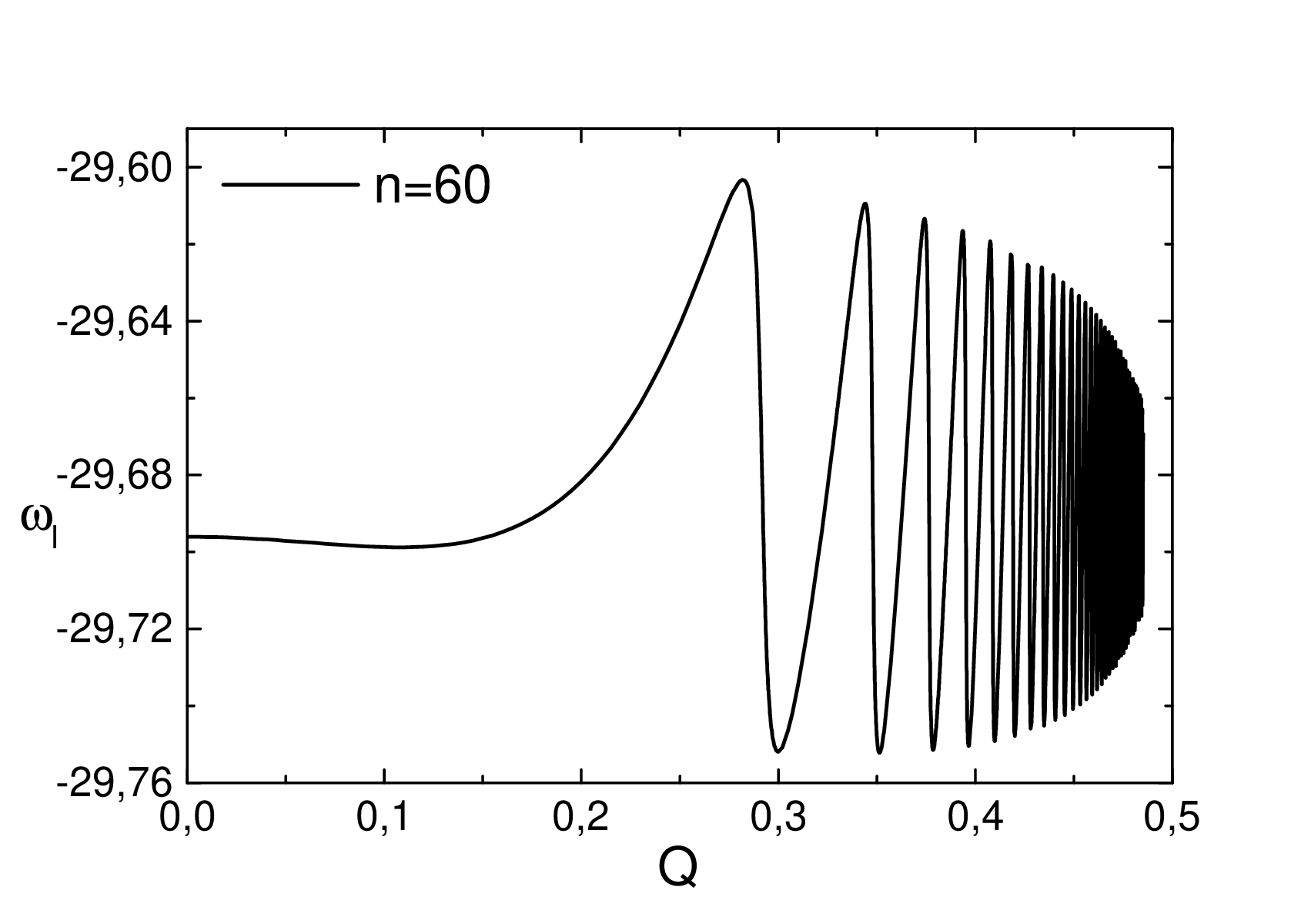}
\includegraphics[angle=270,width=8cm,clip]{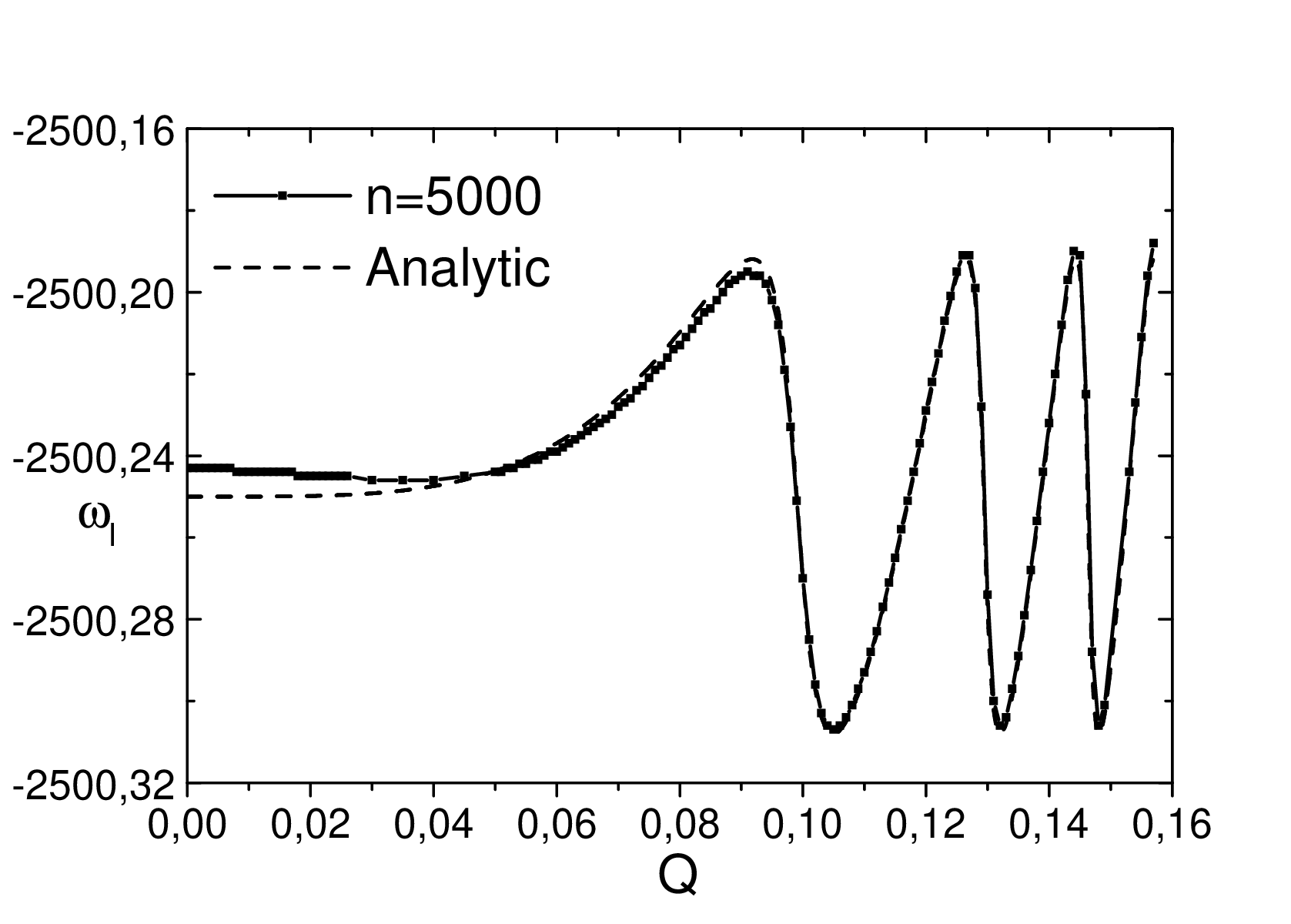}
\caption{
Imaginary part of the RN QNM frequencies as a function of charge for
$n=10,~30,~60,~5000$. For $n=5000$ we also display the actually
computed points, and compare to the prediction of the analytic formula
(\ref{MNf}). As for the real part, the oscillations are reproduced
extremely well, but the disagreement with our numerical data increases
for small charge.
}
\label{fig3}
\end{figure}

Similar considerations apply to the imaginary parts. We present a few
plots illustrating the general trend for the imaginary parts in figure
\ref{fig3}, showing again excellent agreement with the asymptotic
formula (\ref{MNf}) as $n$ increases. Once again, the analytic formula
shows deviations from our ``exact'' numerical results only for small
values of the charge, probably indicating that corrections of order
$n^{-1/2}$ should be taken into account for small values of $Q$. When
we look at the mode trajectories in the complex-$\omega$ plane, the
increasingly oscillating behaviour of the real and imaginary parts
means that the number of ``spirals'' described by the mode before
reaching the extremal value increases roughly as the mode order $n$.

Can we deduce something from the agreement of our numerical results
with formula (\ref{MNf}) at large values of the charge?  It would be
extremely interesting to draw consequences on the extremal RN case,
for various reasons. First of all, the QNM spectrum for extremal RN
black holes is characterized by an isospectrality between
electromagnetic and gravitational perturbations, which has been
motivated in \cite{OOMIKRW} as a manifestation of supersymmetry.
Furthermore, topological arguments have been used to show that the
entropy-area relation breaks down for extreme QNM's
\cite{HHR}. Therefore, we believe that some caution is required in
claiming that the connection between QNM's and the area spectrum is
still valid for extreme black holes, as recently advocated in
\cite{ACL}.  These problems may be connected with our recent finding
that extremally charged black holes in a (non-asymptotically flat)
anti-de Sitter spacetime could be marginally unstable \cite{BK}.

Our numerics seem to indicate that we can trust formula (\ref{MNf}) in
the large-charge regime. Then a very interesting conclusion follows
\cite{N2}: the real part of the frequency for extremal RN black holes
coincides with the Schwarzschild value, i.e.
\be\label{XRN}
\omega_R^{RN}\to {\ln 3\over 4\pi} \qquad \text{as} \qquad Q\to 1/2.
\ee
However, it is quite difficult to check this prediction numerically.
In the extremal RN case, due to the coalescence of the inner and outer
horizons, the singularity structure of the radial perturbation
equations changes. Therefore one has to apply a different (and
slightly more involved) procedure, which has been described in
\cite{OMOI}. We have tried to apply that procedure to get highly
damped QNM's. Apparently, the extremal RN QNM's show a behaviour which
is rather similar to that of Schwarzschild QNM's (see, e.g., figure 1
in \cite{L}): they have finite real part for small values of
$\omega_I$, approach the pure-imaginary axis as $\omega_I$ is
increased, and then the real part increases again. This would support
the predictions of the asymptotic formula (\ref{MNf}). Unfortunately,
we have not yet managed to get stable numerical results for large
values of $\omega_I$. We plan to improve our codes and obtain more
numerically stable results in the future.  

If, supported by the agreement between our numerics and the analytical
prediction, we assume that formula (\ref{XRN}) holds, an interesting
result emerges: Hod's conjecture (\ref{Mhom}) is {\it incompatible}
with the ``truncated'' version of the first law of black hole
mechanics (\ref{Flaw}) obtained {\it dropping the $\Delta Q$ term}
\cite{H}. Indeed, formula (\ref{conj}) predicts that the real part of
the asymptotic frequencies in the extremal case should be zero. This
does not imply that Hod's conjecture is wrong, but only that {\it
dropping the $\Delta Q$ term} to deduce formula (\ref{conj}) is not
a valid assumption.

\subsection{Kerr black holes}

Quasinormal frequencies of Kerr black holes were first studied by
Leaver using continued fraction techniques \cite{L} and subsequently
investigated by other authors \cite{KerrQNM}.  

A systematic exploration of the behaviour of the first few overtones
was carried out only some years ago, using Leaver's continued fraction
method, by Onozawa \cite{O}, who found some rather odd features as the
mode damping increases. For example, Detweiler \cite{De} showed that
the first few modes having $l=m$ have vanishing imaginary part and
real part equal to $m$ as the black hole becomes extremal (in our
units, as $a\to 1/2$). He also showed analytically that there can be
infinite solutions to the Teukolsky equation having $a=1/2$ and
$\omega=m$, which led to the suspicion that {\it all} modes having
$l=m$ should ``cluster'' on the real axis at $\omega=m$ as the black
hole becomes extremal.  An interesting outcome of Onozawa's
investigation was that, for given values of $l$ and $m=l$, there is
at least one mode frequency which does not tend to $m$ in the
extremal limit.  Onozawa also found that, when $n\sim 10$, modes
having negative $m$ (which for the first few overtones show a tendency
to decrease in frequency as the hole is spun up) show instead a
tendency to ``turn around'' and increase their frequency as $a$
increases, sometimes showing strange ``loops'' in the complex-$\omega$
plane (see figures 3 and 4 in \cite{O}).

We have confirmed Onozawa's results, and extended them to moderately
high $n$. However, since one has to solve simultaneously the angular
and the radial continued fraction, the numerical problem turned out to
be much more tricky than the search of highly-damped modes in the RN
case. Comparing Nollert's technique and a standard summation of the
continued fraction using Gautschi's algorithm \cite{L}, we found out
that it is much harder to achieve a stable numerical computation of
modes for $n\sim 50$ or higher \cite{thankCL}. Because of these
convergence problems, even using Nollert's method, we did not manage
to push the numerical calculation to very large values of
$n$. However, even moderately high values of $n$ shed some light on
what should be the asymptotic behaviour of Kerr QNM's.

\begin{figure}[h]
\centering
\includegraphics[angle=270,width=8cm,clip]{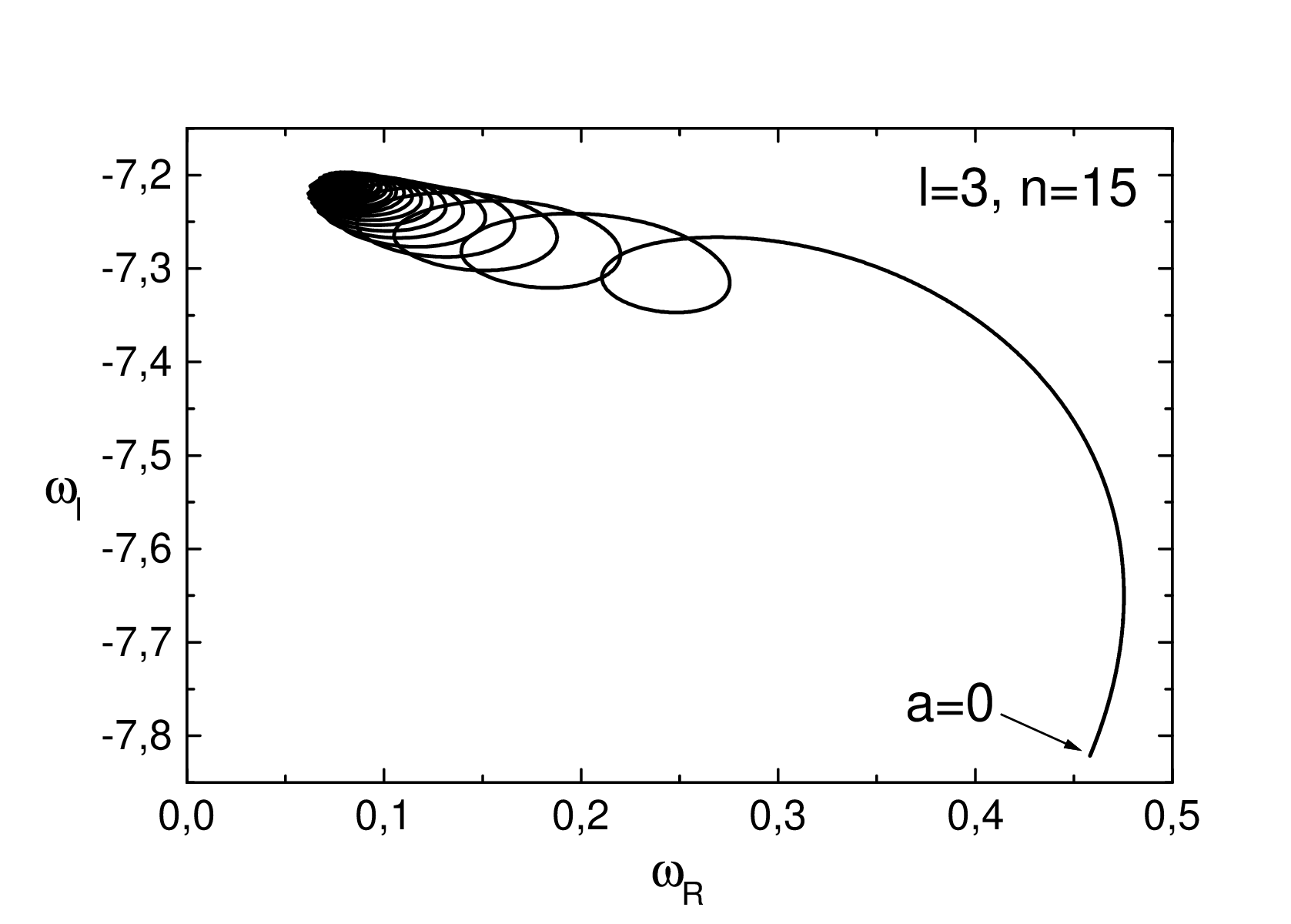}
\includegraphics[angle=270,width=8cm,clip]{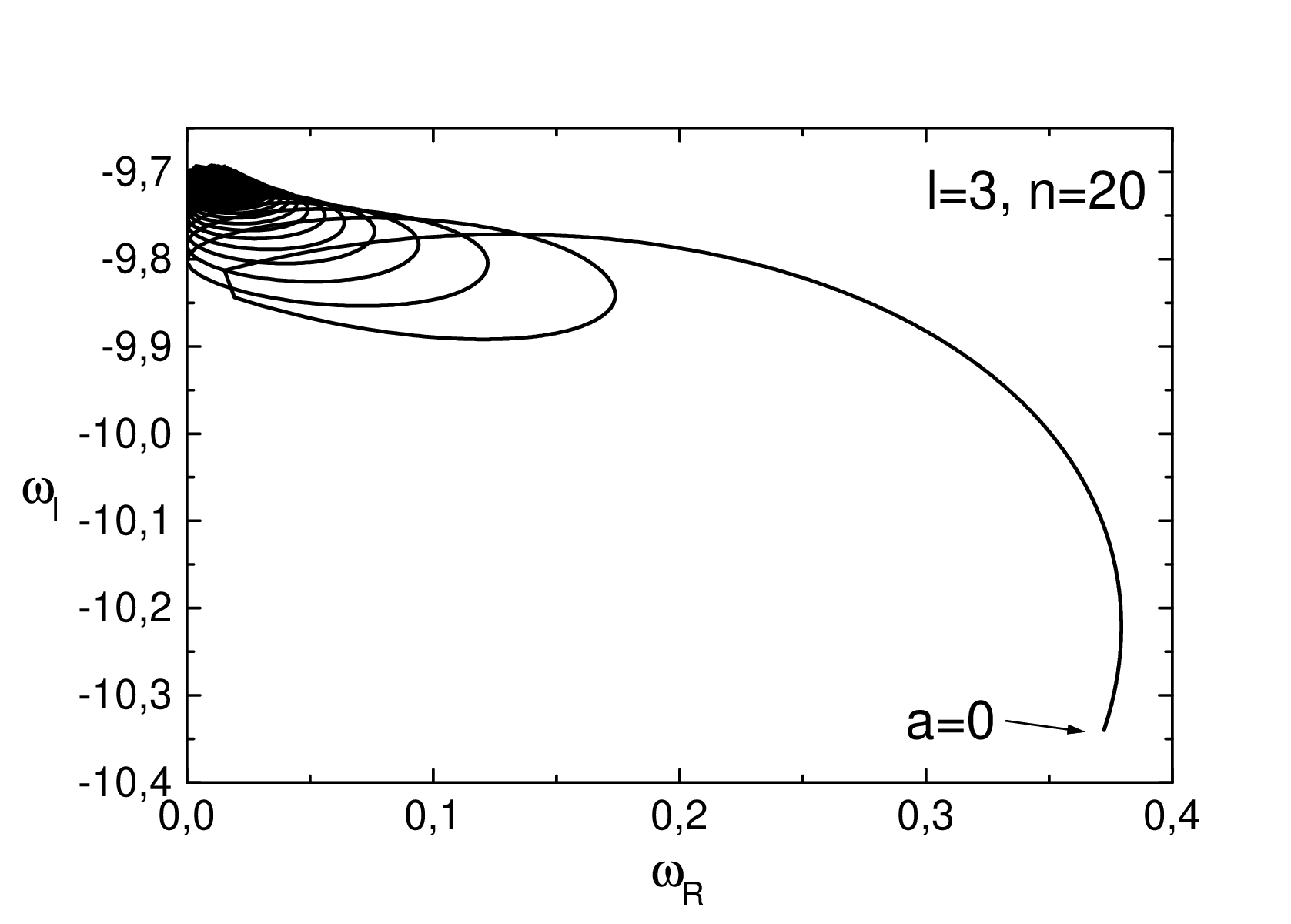}
\caption{
Trajectories of Kerr modes having $m=0$ in the complex-$\omega$ plane.
The left panel corresponds to $l=3,~n=15$ and the right panel to
$l=3,~n=20$. The number of spirals increases with the mode order, as
in the RN case. We have marked by an arrow the point in the plane
corresponding to the Schwarzschild limit.
}
\label{fig4}
\end{figure}

\begin{figure}[h]
\centering
\includegraphics[angle=270,width=8cm,clip]{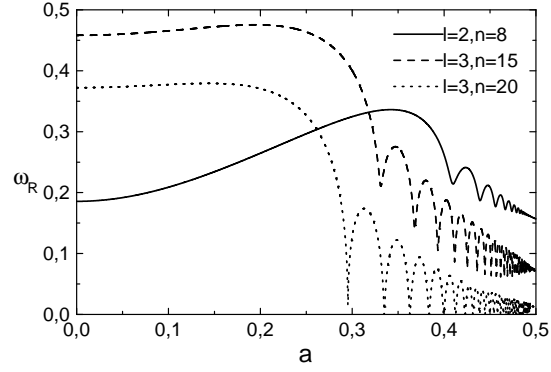}
\caption{
Real part of Kerr modes having $m=0$ as a function of $a$.
Labels indicate the corresponding values of $l$ and of the mode order $n$.
}
\label{fig5}
\end{figure}

\begin{figure}[h]
\centering
\includegraphics[angle=270,width=8cm,clip]{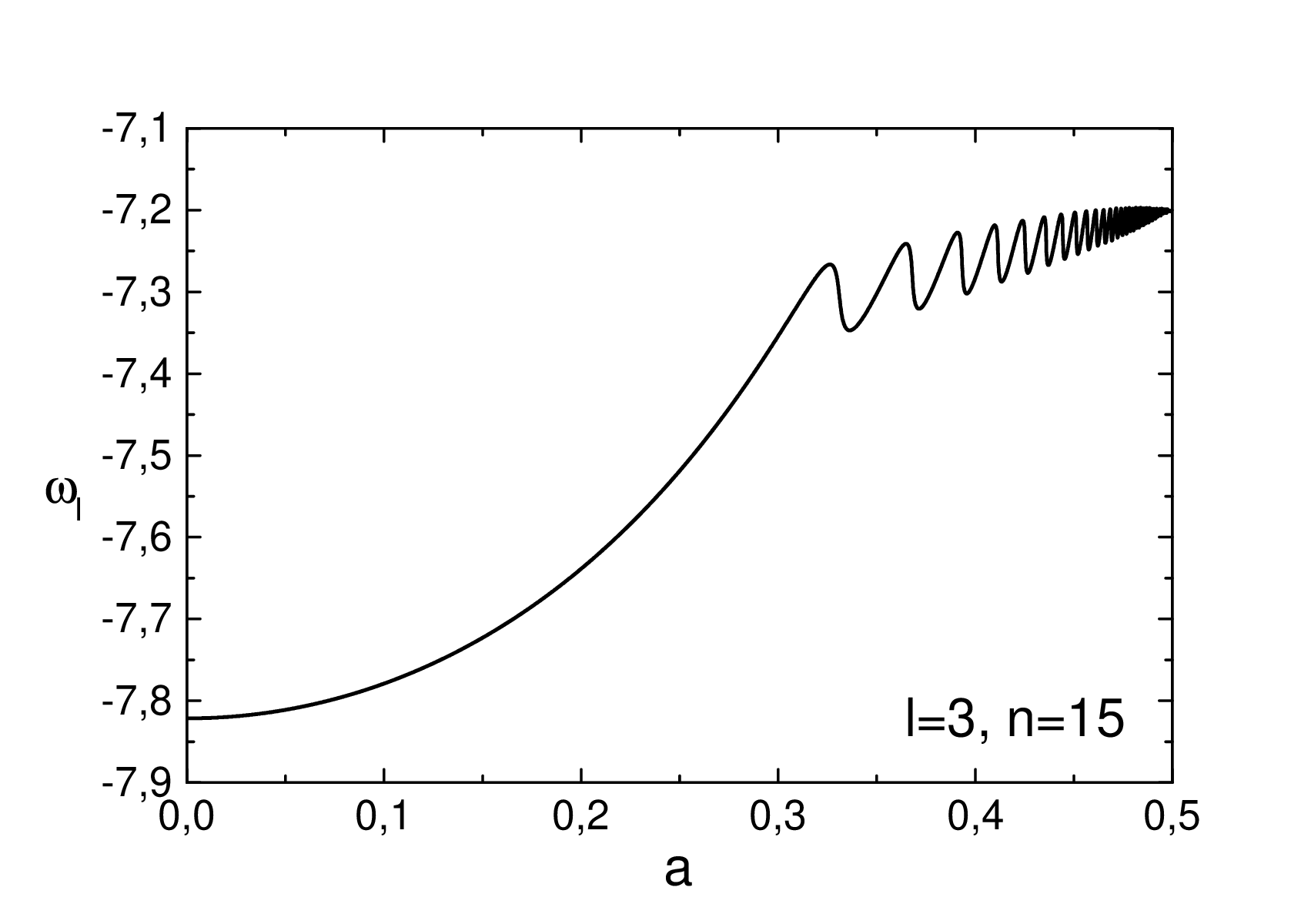}
\includegraphics[angle=270,width=8cm,clip]{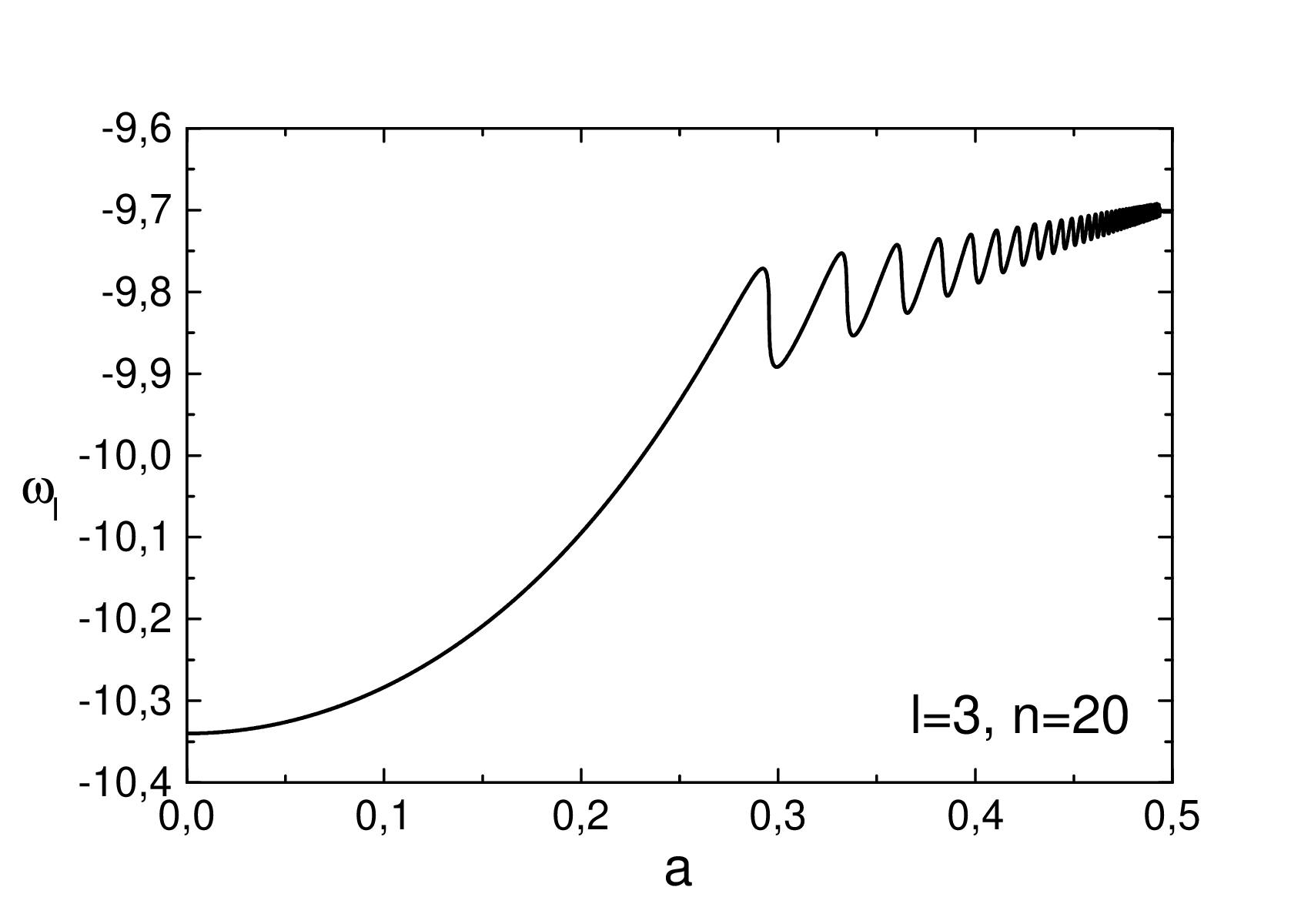}
\caption{
Imaginary part of Kerr modes having $m=0$ as a function of $a$. The
left panel corresponds to $l=3,~n=15$ and the right panel to $l=3,~n=20$.
}
\label{fig6}
\end{figure}

Let us first consider Kerr perturbations having $m=0$. We have been
able to compute quite a few moderately highly damped modes for $l=2$
and $l=3$, and in figure \ref{fig4} we show two of these modes. These
plots should be compared to the RN modes we have shown in figure
\ref{fig1}: there is a similar ``looping'' behaviour, with the number
of loops increasing as the damping of the mode increases. We notice
that a similar looping behaviour has recently been found by
Glampedakis and Andersson for {\it scalar} perturbations of Kerr black
holes, using a different method \cite{GA}. Figure \ref{fig5} and
\ref{fig6} show the real and imaginary parts (respectively) of some
Kerr modes with $m=0$. It is useful to compare these plots with
figures \ref{fig2} and \ref{fig3}: the behaviour is extremely
similar, and asymptotically it can probably be described by some
formula reminiscent of (\ref{MNf}). Of course, such a behaviour is not
even close to that predicted by formula (\ref{conj}).

Hod recently used Onozawa's data, which to our knowledge are the only
available published data for highly damped Kerr modes, to show that
the results predicted by formula (\ref{conj}) agree with the
numerically computations in \cite{O} within $\sim 5\%$, at least when
$l=m$. We repeated Onozawa's calculations, finding excellent agreement
with his results, and then extended it to higher-order modes. We found
that, as $n$ increases, the formula conjectured by Hod does not seem
to provide a good fit to the asymptotic modes. As shown in figure
\ref{fig5}, the proposed formula disagrees quite badly in the small
rotation rate regime even for the low-lying modes which were used for
comparison by Hod. In any event, this is to be expected and does not
contradict Hod's claim, since it is known from previous calculations
in the Schwarzschild limit \cite{L} that modes with $n<10$ have real
parts of the QNM frequencies which are not close enough to their
asymptotic value. However, if formula (\ref{conj}) really holds in
the asymptotic limit, one would expect the agreement with numerical
calculations to get better as the mode's imaginary part increases; on
the contrary, figure \ref{fig4} shows that, as $n$ increases, the
agreement gets worse, even for large rotation rates (which used to
show rather good agreement for $n\sim 10$). So Hod's formula does not
seem to provide an accurate fit to the asymptotic frequencies. We
actually found that, as the mode order increase, modes having $l=m=2$
are fitted extremely well (relative errors being of order $\sim 0.1
\%$ when $n\sim 50$) by the relation
\be\label{mOm} 
\omega^{Kerr}_{l=m=2}=2\Omega+i2\pi T_H n.  
\ee
It is interesting to note that, although the real part tends to a
different asymptotic value, the spacing in the imaginary parts {\it
does} agree with the value conjectured by Hod in \cite{H2}.  Indeed,
Hod put forward his conjecture observing that in the Schwarzschild
case the asymptotic QNM spacing is given by $2\pi T_H$. The
convergence to the indicated asymptotic behaviour is faster for large
values of $a$, and generally the agreement between formula (\ref{mOm})
and our numerics is good for values of $a\gtrsim 0.1$.  The index $n$
appearing in formula (\ref{mOm}) depends, of course, on the labelling
convention we use to ``count'' QNM's.

Notice also that the modes' imaginary part does not show the typical
``shift'' $(-1/4)$ which is present in formula (\ref{Mresult}). In our
opinion, this is another hint that the Schwarzschild limit of highly
damped Kerr modes should be taken with special care, and that order of
limits issues may be relevant in the asymptotic regime.

Preliminary calculations show that modes having $l=2$ and $m=1$,
$m=-1$ and $m=-2$ show a more complicated behaviour. We are currently
trying to improve our understanding of highly damped Kerr QNM's using
both analytical and numerical techniques. Some of our results will be
shown in a separate paper \cite{BCKO}.

\section{Conclusions}

In this paper we have numerically investigated the asymptotic
behaviour of QNM's for charged (RN) and rotating (Kerr) black holes.
We have first confirmed Nollert's results and extended them to scalar
perturbations of Schwarzschild black holes. Our numerics are
consistent with
\be
\omega_R={\ln 3\over 4\pi}+{k_s[(s^2-1)-3l(l+1)] \over \sqrt{n}},
\ee
where, within our accuracy, $k_{0}=-0.0970=3k_{-2}$. Recently the
constant $k_{-2}$ has been determined analytically - see formula
(\ref{km2}) - and our result may be useful to determine $k_0$ as well.

More importantly, our results for charged and rotating black holes do
not agree with the simple behaviour predicted by Hod's conjecture for
the real part of the frequency, as given in formula (\ref{conj}).  We
have shown that both the real and the imaginary part of RN QNM's as
functions of charge display an oscillating behaviour. The oscillations
start at smaller values of the charge and get faster as $n$
increases. We have compared our numerical results to the predictions
of the asymptotic formula (\ref{MNf}) derived in \cite{MN} and found
that, in general, they agree extremely well, especially for large
values of the mode index $n$. The formula derived by Motl and Neitzke
only fails to reproduce our numerical results in the limit of small
charge.  Quite likely, their expression has a wrong limit in the
Schwarzschild case (the real part tending to $\ln 5/4\pi$ instead of
$\ln 3/4\pi$) because finite-$n$ corrections become relevant as $Q$
tends to zero \cite{N2}. A computation of higher-order corrections to
the asymptotic RN formula will probably give a final answer on the
reasons for the small-$Q$ discrepancy we observe. If we trust the
predictions of formula (\ref{MNf}), as the agreement between
analytical and numerical results suggests to do, the asymptotic
frequency for extremal RN black holes is the same as for Schwarzschild
black holes. This implies that dropping the $\Delta Q$ term in the
first law of black hole mechanics, as done by Hod to conjecture the
validity of formula (\ref{conj}), is presumably {\it not} a justified
assumption.

Highly damped gravitational quasinormal frequencies of rotating (Kerr)
black holes proved more difficult to compute, even using Nollert's
method, and show a more complicated behaviour. For $m=0$, we observe
spirals in the complex plane reminiscent of RN modes, confirming the
behaviour found in \cite{GA} for highly damped {\it scalar}
perturbations. For $l=m=2$, the behaviour is completely different. The
asymptotic behaviour of Kerr modes having $l=m=2$ is very well fitted
by formula (\ref{mOm}), that we rewrite here:
\be\nn
\omega^{Kerr}_{l=m=2}=2\Omega+i2\pi T_H n,
\ee 
where $\Omega$ is the angular velocity of the black hole horizon and
$T_H$ its temperature. The convergence to the limiting value is faster
when $a$ is large, and the formula can be see as an extension of
formula (\ref{Mresult}) to Kerr modes having $l=m=2$. We think that a
calculation of finite-$n$ corrections to the asymptotics may help
explain both the faster convergence rate at large $a$, and the
apparent disagreement with formula (\ref{Mresult}) in the limit $a\to
0$. A more extensive investigation of asymptotic Kerr QNM's is ongoing
\cite{BCKO}.

\acknowledgments

We are grateful to N. Andersson for useful discussions and a careful
reading of a first draft of the manuscript. It is a pleasure to thank
A. Neitzke for e-mail correspondence, and especially for sharing with
us his unpublished results on the predictions of the analytic RN
formula. Finally, we thank A. Ashtekar and J. Pullin for their
encouragement to carry out the present calculation. This work has
been supported by the EU Programme 'Improving the Human Research
Potential and the Socio-Economic Knowledge Base' (Research Training
Network Contract HPRN-CT-2000-00137).

\begin{figure}[h]
\centering
\includegraphics[angle=270,width=8cm,clip]{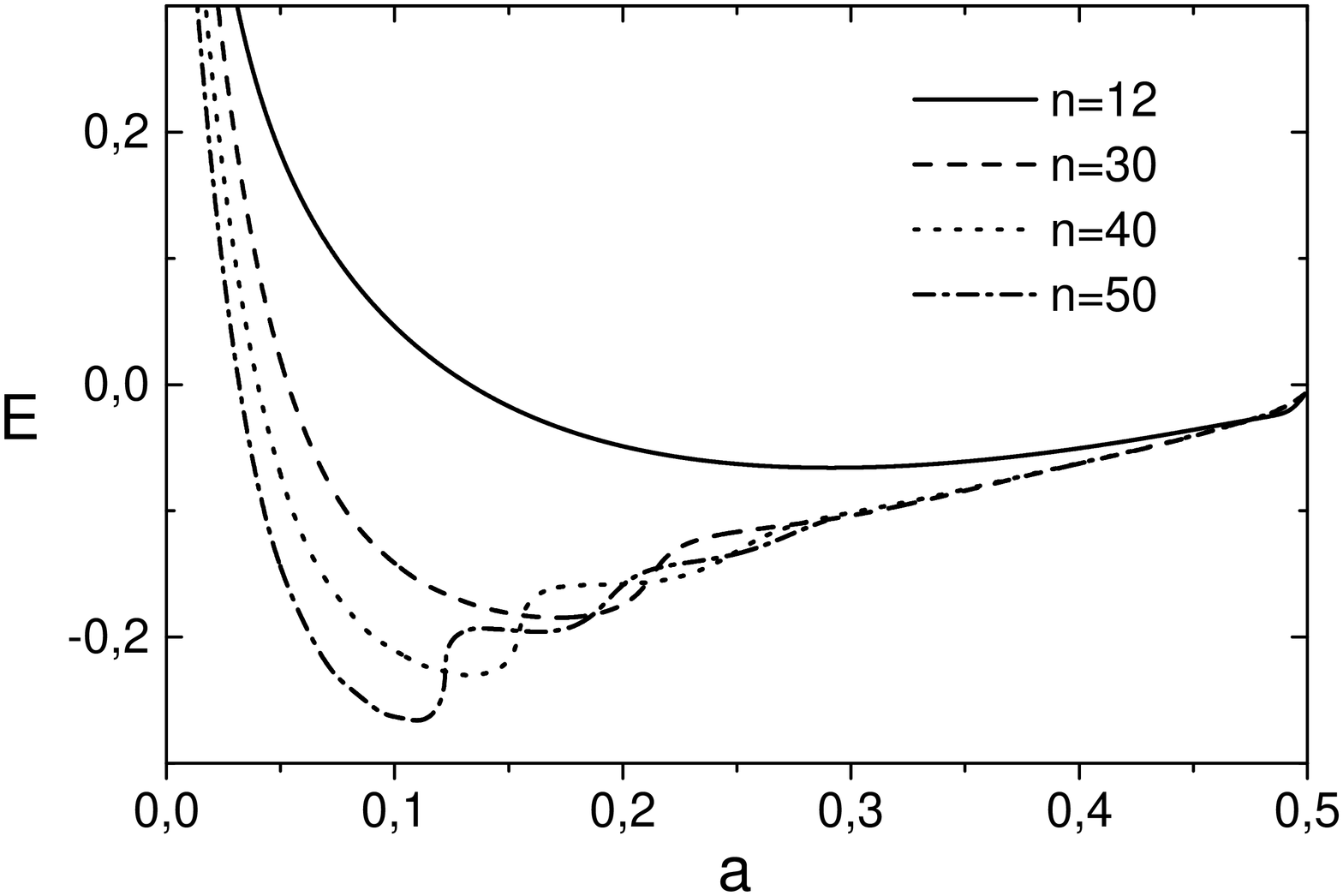}
\caption{
Relative error in the formula for the asymptotic frequency conjectured
by Hod. The plots shows $E=(\omega_R-\tilde \omega_R)/\tilde
\omega_R$, where $\tilde \omega_R$ is defined in formula (\ref{conj}),
for increasing values of the mode index $n$, namely $n=12,~30,~40,~50$
and $l=m=2$. As $n$ grows, the relative error in the conjectured
asymptotic formula tends to be larger and larger.
}
\label{fig7}
\end{figure}


\begin{references}
\bibitem{units} 
In our discussion of the numerical results we consistently use
Leaver's convention on units, setting $2M=1$; in particular, this
means that extremal Kerr and RN black holes correspond, respectively,
to $a=1/2$ and $Q=1/2$.
\end{references}
\end{document}